\documentclass[preprint,12pt, a4paper]{elsarticle}
\biboptions{sort&compress}

\usepackage{amssymb}
\usepackage{lineno}
\usepackage{float}
\restylefloat{table}

\usepackage[utf8]{inputenc}
\usepackage[T1]{fontenc}
\usepackage[american]{babel}
\usepackage{amsmath}
\usepackage{booktabs}
\usepackage{xcolor}
\usepackage{siunitx}
\DeclareSIUnit\angstrom{\text{\smash[t]{\AA}}}
\usepackage[version=4]{mhchem}
\usepackage{listings}
\usepackage{xurl}
\usepackage[destlabel]{hyperref}
\definecolor{linkblue}{RGB}{0,70,120}
\hypersetup{
  colorlinks=true,
  allcolors=linkblue,
  pdfborder={0 0 0},
  linktoc=all,
  bookmarksnumbered=true,
  bookmarksdepth=2
}
\newcommand*{\suppref}[2]{#2}
\journal{SoftwareX}

\begin{document}

\begin{frontmatter}

\title{AmorphGen: A Python package for amorphous structure generation
  with machine-learning and classical interatomic potentials}

\author[label1]{Chaiyawat Kaewmeechai\corref{cor1}}
\ead{c.kaewmeechai@bham.ac.uk}
\author[label1]{Louie Slocombe}
\author[label1]{Alexis G. Manche}
\author[label1]{David O. Scanlon\corref{cor1}}
\ead{d.o.scanlon@bham.ac.uk}
\cortext[cor1]{Corresponding authors}
\address[label1]{School of Chemistry, University of Birmingham, Edgbaston, Birmingham B15 2TT, United Kingdom}

\begin{abstract}
AmorphGen is an open-source Python package for generating and analyzing amorphous structures from a composition or a crystal. Chemically informed rules estimate initial densities and assign separation and coordination constraints. Three routes are available: random placement followed by relaxation, melt-quench from a crystal, and a hybrid route that anneals and quenches random structures. ASE provides access to classical and machine-learning potentials, with GPU batching for selected backends. Analysis tools compute coordination numbers, bond angles, rings, and
X-ray or neutron structure factors for comparison with measurements. An optional interface to the FEFF code computes extended X-ray absorption fine structure (EXAFS) spectra. For amorphous Ga$_2$O$_3$, TiO$_2$ and InGaZnO$_4$, the cation coordination numbers are close to reference models and measurements, and the X-ray structure factors reproduce the main measured peaks. The EXAFS spectra for Ga$_2$O$_3$ and InGaZnO$_4$ also capture the dominant first-shell features observed experimentally. The hybrid route reduces bonding
defects in SiO$_2$ and As and builds the first sharp diffraction peak of
SiO$_2$. By simplifying ensemble generation and standardizing structural
assessment, AmorphGen makes it practical to compare amorphous models
across compositions, potentials, and preparation routes, and to screen
new amorphous compositions.
\end{abstract}


\end{frontmatter}

\section*{Metadata}
\label{metadata}

\begin{table}[H]
\begin{tabular}{|l|p{5.9cm}|p{5.9cm}|}
\hline
Nr. & Code metadata description & \\
\hline
C1 & Current code version & v1.0.0  \\
\hline
C2 & Permanent link to code/repository used for this code version & \url{https://github.com/SMTG-Bham/AmorphGen} \\
\hline
C3 & Legal code license & MIT License \\
\hline
C4 & Code versioning system used & git \\
\hline
C5 & Software code languages, tools, and services used & Python 3.10 to 3.14; ASE, NumPy, SciPy, PyYAML, Matplotlib; optional PyTorch through MACE, CHGNet, SevenNet, and TorchSim; optional pyace and LAMMPS; optional external FEFF8.5L for EXAFS spectra; GitHub Actions \\
\hline
C6 & Compilation requirements, operating environments, \& dependencies & Linux or macOS; CUDA GPU optional (CPU and Apple MPS supported); \texttt{pip install amorphgen[mace]}, \texttt{[chgnet]}, or \texttt{[sevennet]}; the TorchSim engine needs the \texttt{[torchsim]} extra, Python 3.12 or later and a CUDA GPU or CPU (not Apple MPS) \\
\hline
C7 & Link to developer documentation/manual (if available) & \url{https://smtg-bham.github.io/AmorphGen} \\
\hline
C8 & Support email for questions & \href{mailto:c.kaewmeechai@bham.ac.uk}{\nolinkurl{c.kaewmeechai@bham.ac.uk}} \\
\hline
\end{tabular}
\end{table}


\section{Motivation and significance}
\label{motivation}

Amorphous materials are widely used in thin-film transistors,
solid-state electrolytes and phase-change memory, and simulations of
their electronic and ionic properties start from atomistic models. Random
structure searching (RSS) places atoms under distance constraints and
optimizes the structures \cite{pickard2011,youn2014,jones2020glassy},
whereas melt-quench
molecular dynamics (MQ-MD) cools a simulated liquid into a glass
\cite{stich1991amorphous}. For RSS, each composition needs a starting
density, minimum separations, and, for ionic materials, coordination
targets. Inappropriate values lead to bonding defects,
such as anion--anion dimers or metal--metal bonds in a stoichiometric
oxide.

Most existing tools leave starting parameters and much of the preparation
workflow to the user. The physically informed seed-coordinate method
\cite{youn2014}, Packmol \cite{martinez2009packmol}, and AIRSS
\cite{pickard2011} take separations and density as inputs. Bond switching
builds high-quality continuous random networks but requires a bonding
network and force field for each system
\cite{wooten1985,barkema2000crn,zheng2025gbsmc}. Reverse Monte Carlo codes
\cite{mcgreevy1988rmc,tucker2007rmcprofile,aoun2016fullrmc,opletal2008hrmc}
and the gradient-based TorchDisorder \cite{gore2026torchdisorder} fit
atomic structures to measured structure factors and therefore require
experimental data for the composition of interest.
LAMMPS \cite{thompson2022lammps} requires each simulation stage to be
scripted, while Atomsk \cite{hirel2015atomsk} and pyiron
\cite{janssen2019} provide building blocks and workflow management
without amorphous-specific defaults.
Vitriflow \cite{cottom2026vitriflow} calibrates melt-quench ensembles with
LAMMPS from a user-defined random cell. amorphouspy \cite{amorphouspy}
builds random oxide glasses from a composition, with the density from an
empirical model, and quenches them with classical force fields.

More affordable interatomic potentials make larger MQ-MD simulations
possible, but reliable dynamics still depend on the potential and
protocol. \textit{Ab initio} MD is computationally expensive, which restricts cell
sizes and forces cooling rates far above experiment. Workflows such as
MPmorph \cite{aykol2018mpmorph} automate it, but start from random
structures packed with Packmol \cite{martinez2009packmol}. Classical force fields enable thousands of
atoms but are fitted to particular systems
\cite{vanbeest1990bks,carre2008chik,oeffner1998}. Foundation
machine-learning interatomic potentials (MLIPs), including MACE
\cite{batatia2022mace,batatia2023foundation}, CHGNet
\cite{deng2023chgnet}, and SevenNet \cite{park2024sevennet}, provide forces
across most of the periodic table without system-specific fitting,
allowing thousands of atoms and nanosecond trajectories. Static accuracy does not ensure stable finite-temperature dynamics
\cite{li2025finitet}: under a barostat at high temperature, foundation
MLIPs can expand amorphous oxides far below their real density
\cite{li2026meltquench}. 
Amorphous structures remain a weak point of foundation MLIPs
\cite{fragapane2026amorphous}. Fine-tuning needs less reference data than
training a new potential \cite{fragapane2026amorphous,tompa2026finetuning};
training from scratch \cite{deringer2018gap,deringer2019adma}, automated
by tools such as autoplex \cite{liu2025autoplex}, requires a new potential
for each chemical system. A preparation protocol and checks against
reference data remain essential, whichever potential is used.

The missing step is to connect composition-derived starting parameters,
ensemble preparation suited to MLIPs, and comparison with reference data
in one workflow. Existing tools either analyze structures
\cite{leroux2010rings,stukowski2010ovito} or automate the preparation of
amorphous ensembles \cite{aykol2018mpmorph,cottom2026vitriflow,amorphouspy}. To our
knowledge, however, no package takes a user from a chemical formula to
amorphous structures with minimum separations and coordination targets
derived automatically from the chemical bonding of the composition.

\begin{figure}[p]
\centering
\includegraphics[width=0.9\columnwidth]{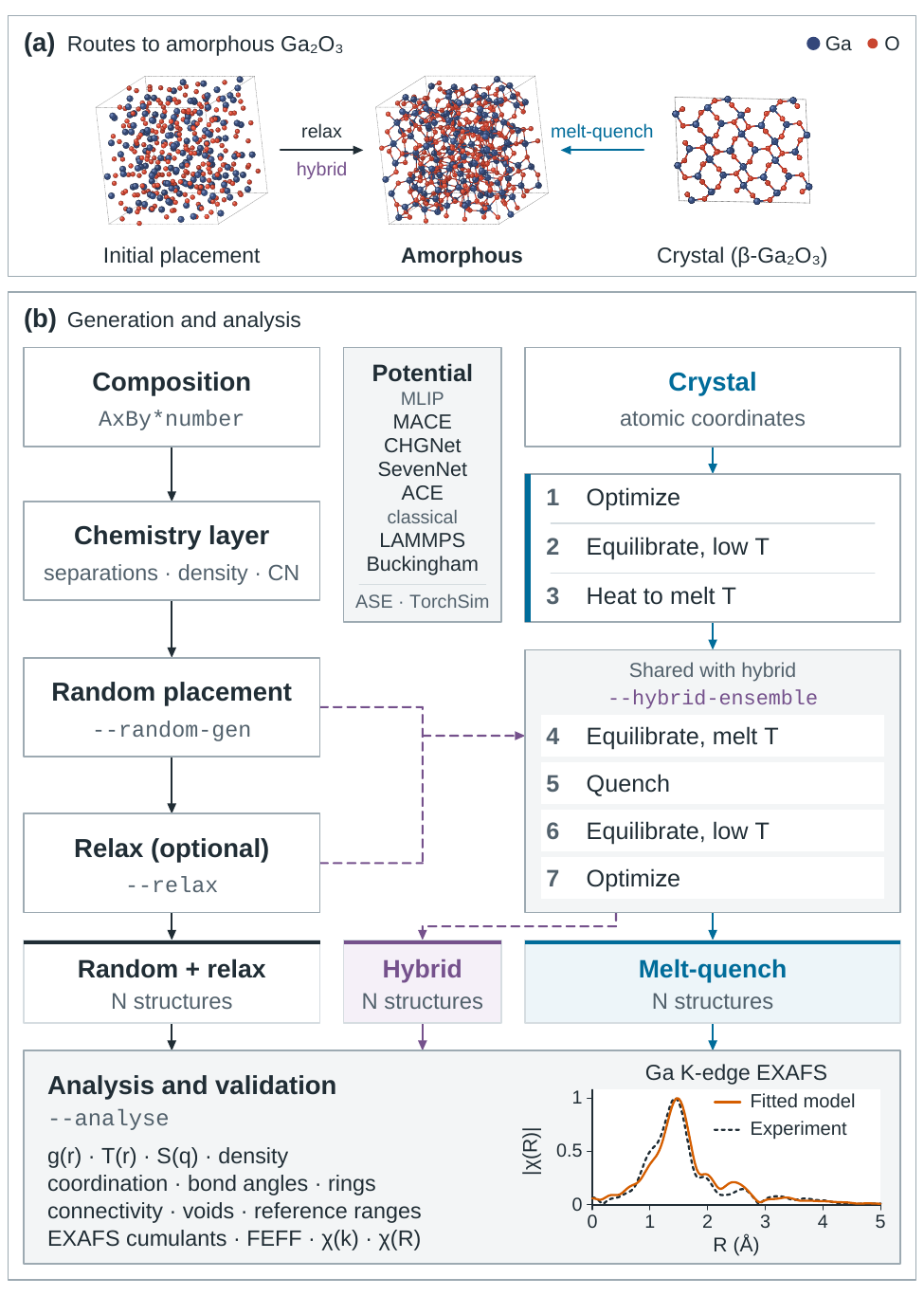}
\caption{Overview of AmorphGen. (a) An amorphous \ce{Ga2O3} structure
(center, made by the hybrid route), reached from a random initial
placement (left) by relaxation or the hybrid route, or from the
$\beta$-\ce{Ga2O3} crystal (right) by melt-quench.
(b) Generation and analysis: the random + relax, hybrid, and melt-quench routes,
the machine-learning or classical potentials that drive them (center),
and the analysis tools. The plot shows the Ga K-edge EXAFS of the 100-cell a-\ce{Ga2O3}
random + relax ensemble against the film of Reis \textit{et al.}\
\cite{reis2023}, fitted as in Figure~\ref{fig:exafs}a.}
\label{fig:overview}
\end{figure}

Here, we present AmorphGen, an open-source Python library and command-line
tool that derives a starting density, minimum separations, and
coordination targets from composition across 21 material classes, then automates ensemble
preparation and structural analysis (Figure~\ref{fig:overview}). It offers three routes: random + relax, melt-quench from a crystal, and a
hybrid route that anneals and quenches random structures. It accepts any
interatomic potential with an Atomic Simulation Environment (ASE)
calculator \cite{larsen2017}. We compare the starting-density estimate
with a 100-compound calibration set and test the random + relax and hybrid routes on five
materials: \ce{Ga2O3}, \ce{TiO2}, \ce{InGaZnO4}, \ce{SiO2}, and As.
The ensembles are compared with reference structures, measured X-ray
structure factors, and, for \ce{Ga2O3} and \ce{InGaZnO4}, extended X-ray absorption fine
structure (EXAFS) spectra calculated through an interface to FEFF
\cite{ankudinov1998feff8,rehr2000rmp}.

\section{Software description}
\label{description}

\subsection{Software architecture}
\label{sec:architecture}

AmorphGen is a pure-Python package built on ASE \cite{larsen2017},
NumPy \cite{harris2020numpy}, SciPy \cite{virtanen2020scipy}, PyYAML
\cite{pyyaml}, and Matplotlib \cite{hunter2007}; PyTorch
\cite{paszke2019pytorch} is optional. Extras add MACE, CHGNet,
SevenNet, pacemaker ACE potentials
\cite{drautz2019ace,lysogorskiy2021pace,bochkarev2022pacemaker},
LAMMPS pair styles through ASE's \texttt{LAMMPSlib}
\cite{thompson2022lammps}, and TorchSim \cite{torchsim2025}.
TorchSim batches relaxation and fixed-volume MD for MACE, SevenNet,
Lennard-Jones, and Buckingham--Coulomb potentials.

Four modules organize the workflow (Figure~\ref{fig:overview}):
\begin{itemize}
\item \texttt{amorphgen.pipeline}: random generation and MD protocols;
\item \texttt{amorphgen.utils}: composition rules and interfaces to calculators and engines;
\item \texttt{amorphgen.analysis}: structural descriptors, X-ray and neutron
  structure factors \cite{waasmaier1995,sears1992}, EXAFS, energy ranking,
  and reference comparisons;
\item \texttt{amorphgen.configs}: default protocols and the YAML schema.
\end{itemize}

The \texttt{amorphgen} command accepts flags or schema-validated YAML;
\texttt{amorphgen-slurm} writes Slurm job scripts \cite{yoo2003slurm}.
MD checkpoints and extended-XYZ trajectories with momenta allow
resumption from the last complete frame. Divergence guards stop
unstable MD or relaxation
(\suppref{sec:provenance}{Supplementary Section~S3}), and NumPy's
\texttt{SeedSequence} derives ensemble seeds from one global seed.

\subsection{Software functionalities}
\label{sec:functionalities}

A formula and a multiplier specify the composition. For example,
\texttt{Ga2O3*80} gives a 400-atom cell of 80 formula units with
\ce{Ga2O3} stoichiometry. Charge balance assigns
oxidation states and selects one of 21 material classes. Minimum
separations use Shannon ionic \cite{shannon1976}, Cordero covalent
\cite{cordero2008}, or Goldschmidt metallic \cite{greenwood1997} radii,
with bond-dependent scale factors of 0.70--0.85 and separate
cation--cation rules
(\suppref{sec:composition}{Supplementary Section~S1}).
The class packing fraction $f_{\text{pack}}$ sets the cell:
\begin{equation}
V_{\text{cell}} = \frac{1}{f_{\text{pack}}}\sum_i N_i\,\frac{4}{3}\pi r_i^3,
\qquad
\rho = \frac{\sum_i N_i M_i}{N_{\mathrm{A}}\,V_{\text{cell}}},
\qquad
L = V_{\text{cell}}^{1/3},
\label{eq:cell}
\end{equation}
where $N_i$ and $M_i$ are the atom count and molar mass of element $i$.
This estimate is used as the starting density when no target density is given. Against 100 target densities,
41 measured on amorphous samples and 59 assumed to be 0.80 times the crystal
density, it has a mean absolute relative error of \SI{11.9}{\percent}
(\suppref{sec:packing-calibration}{Supplementary Section~S2}).
These derived values can be set by the user.

The random generator places atoms one at a time, following the
seed-coordinate scheme \cite{youn2014}. Each new atom goes between the
minimum separation and the bonding cutoff from an existing atom that has
not yet reached its coordination target.
The detailed rules are in
\suppref{sec:composition}{Supplementary Section~S1}.

The seven-stage melt-quench pipeline starts from a crystal
(Figure~\ref{fig:overview}). The hybrid route applies stages 4--7
to random structures before or after relaxation: liquid annealing,
quenching, low-temperature equilibration, and final optimization.
Melt-quench ensembles and batch quenching apply stages 5--7 to liquid
snapshots or files in a directory. Stage ensembles, temperatures,
durations, and restart points are configurable; defaults are in
\suppref{sec:provenance}{Supplementary Section~S3}.
TorchSim MD holds the input cell fixed. ASE follows the stage
settings, so the hybrid examples with foundation potentials need
\texttt{-{}-eq-high-ensemble NVT}. Foundation MLIPs can expand the cell under a barostat at high
temperature \cite{li2026meltquench}, so our melt-quench example also
fixes the volume during heating and quenching.

The analysis module computes geometric first-shell cumulants and
single-Gaussian fit bias. Its FEFF interface
\cite{ankudinov1998feff8,rehr2000rmp} builds absorber-centered clusters,
runs an external executable in parallel, checks outputs, and averages
$\chi(k)$ first over sites, then structures, before Fourier
transformation. Reference-spectrum validation and path-truncation
checks are detailed in
\suppref{sec:exafs-checks}{Supplementary Sections~S6.2} and
\suppref{sec:exafs-convergence}{S6.4}.

\subsection{Sample code snippets and analysis}
\label{sec:code}
These commands generate and relax 100 400-atom a-\ce{Ga2O3} cells,
anneal 20, and analyze their structures and Ga K-edge EXAFS.
Steps 1--2 require the \texttt{[mace]} and \texttt{[torchsim]} extras,
Python 3.12 or later, and a CUDA GPU; step~3 reads a reference file from
the \texttt{examples/} directory of the repository, which is not
installed by \texttt{pip}; step~4 requires external FEFF.
Placement stays at \SI{4.84}{\gram\per\cubic\centi\metre} with
\texttt{-{}-retry-mode none}; MACE-MPA-0 then relaxes the atoms and
cubic cell to \SI{0.05}{\electronvolt\per\angstrom}, ten cells at a time.
The hybrid MD uses single precision and fixed volume; its thermal
schedule and the reference-file checks are specified in
\suppref{sec:provenance}{Supplementary Section~S3}.

\begin{lstlisting}
# 1. Random placement and relaxation of 100 Ga160O240 cells
amorphgen --random-gen --composition "Ga2O3*80" -n 100 \
    --seed 2026 --target-density 4.84 --retry-mode none \
    --relax -m mace-mpa-0 --engine torchsim --batch-size 10 \
    -f 0.05 -o random_ga2o3/
# 2. Hybrid route on 20 of them (50 ps anneal at 3000 K)
mkdir -p seeds_ga2o3
cp $(ls random_ga2o3/random_opt/*.xyz | head -n 20) seeds_ga2o3/
amorphgen --hybrid-ensemble --input-dir seeds_ga2o3/ \
    --seed 2026 -m mace-mpa-0 --dtype float32 \
    --engine torchsim --batch-size 10 --eq-high-ensemble NVT \
    --eq-high-steps 100000 -f 0.05 -o hybrid_ga2o3/
# 3. Structural analysis, checked against the reference ranges
amorphgen --analyse --input-dir hybrid_ga2o3/final/ \
    --reference examples/reference_a_Ga2O3.yaml --sq \
    --connectivity --rings --ring-criterion primitive \
    --save-plot plots/
# 4. Ga-O cumulants and Ga K-edge EXAFS over all Ga sites
amorphgen --analyse --input-dir hybrid_ga2o3/final/ \
    --exafs Ga-O --harmonic-fit Ga-O --feff-write feff \
    --feff-absorber Ga --feff-radius 10 \
    --feff-stoichiometry structure --feff-card RPATH=8 \
    --feff-card "CRITERIA=1.0 0.5" --feff-run --feff-workers 8 \
    --feff-kmax 12 --save-plot exafs/
\end{lstlisting}

For a crystal, this command heats and holds the melt at fixed volume
before generating up to 20 quenches, with NPT equilibration at
\SI{300}{\kelvin}.

\begin{lstlisting}
amorphgen Ga2O3_supercell.xyz --mq-ensemble -n 20 \
    -m mace-mpa-0 --melt-ensemble NVT --eq-high-ensemble NVT \
    --eq-high-steps 100000 --eq-low-ensemble NPT -o mq_ga2o3/
\end{lstlisting}

The Python API also supports placement and analysis without relaxation:

\begin{lstlisting}[language=Python]
from amorphgen import generate_random
from amorphgen.analysis import StructureAnalyser

cells = [generate_random({"Ga": 160, "O": 240}, seed=s,
                         target_density=4.84, retry_mode="none")
         for s in range(3)]
sa = StructureAnalyser(cells)
print(sa.coordination("Ga-O")["Ga-O"]["mean"])  # mean Ga-O CN
print(sa.dimer_report()["total"])                # close contacts
\end{lstlisting}

\section{Illustrative examples}
\label{examples}
We tested AmorphGen on five systems: three metal oxides (a-\ce{Ga2O3},
a-\ce{InGaZnO4}, and a-\ce{TiO2}), silica glass (a-\ce{SiO2}), and
amorphous As. For a-As, a MACE potential
trained on r$^2$SCAN data and a model structure were recently published
\cite{liu2026as}, together with the measured structure factor of
Ref.~\cite{smith1975} in digitized form, which allows a direct comparison. Every system was run with the foundation
MLIP MACE-MPA-0 \cite{batatia2023foundation}, trained on the
MPtrj \cite{deng2023chgnet} and sAlex
\cite{schmidt2023adma,schmidt2024alexandria,barrosoluque2024omat24} datasets. For a-\ce{SiO2}, we also used the classical
potential of van Beest, Kramer, and van Santen (BKS)
\cite{vanbeest1990bks} and, for a-As, the potential of Liu \textit{et al.},
loaded through \texttt{-{}-model-path}.

For each oxide, 100 structures were made by random placement and
relaxation and 20 by the hybrid route, following the workflow in
Section~\ref{sec:code} with that oxide's composition and density. The engines,
precision, and per-task seeds are given in \suppref{sec:provenance}{Supplementary Section~S3}
and \suppref{tab:ensemble-provenance}{Table~S4}. The cubic cells
hold 400~atoms for a-\ce{Ga2O3}, 350 for a-\ce{InGaZnO4}, 270 for
a-\ce{TiO2}, and 648 for a-\ce{SiO2}.
For a-As, five structures were made by each route in 2000-atom cells,
the size of the model of Liu \textit{et al.}
Every cell was built at a supplied reference density (for
a-\ce{Ga2O3}, the PBE0-TC-LRC value \cite{kaewmeechai2025}), so these
tests check the placement rules and the protocol, not the density
estimate.

\begin{figure}[t]
\centering
\includegraphics[width=0.98\textwidth]{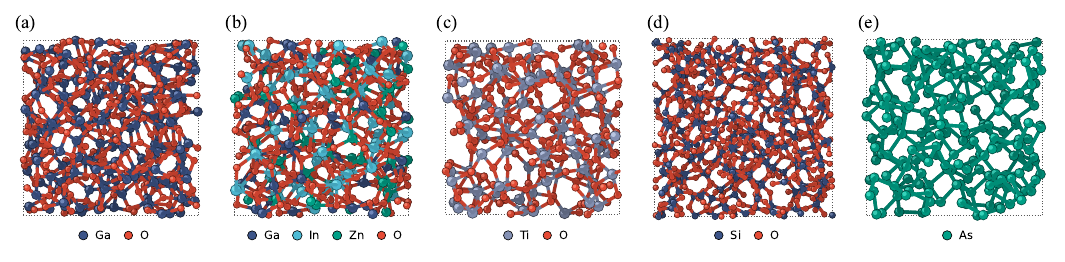}
\caption{One relaxed structure from the random + relax ensembles of (a)
a-\ce{Ga2O3} (400 atoms), (b) a-\ce{InGaZnO4} (350), (c) a-\ce{TiO2}
(270), (d) a-\ce{SiO2} (648), and (e) a-As, projected on the $x$--$y$
plane, with cation--O and As--As bonds. For clarity, (e) shows a
216-atom a-As cell made with the same settings as the 2000-atom cells
analyzed in Section~\ref{examples}.}
\label{fig:structures}
\end{figure}

Below, the \emph{anneal} is the \SI{50}{\pico\second} fixed-volume stage of the
hybrid route (\SI{1500}{\kelvin} for a-As and \SI{3000}{\kelvin} for the oxides). Each GPU
calculation ran on one NVIDIA A100-SXM4 GPU with four cores of an
Intel Xeon Ice Lake CPU, and the BKS runs used four CPU cores. Figure~\ref{fig:structures} shows one structure per system.

\begin{figure}[H]
\centering
\includegraphics[width=0.8\textwidth]{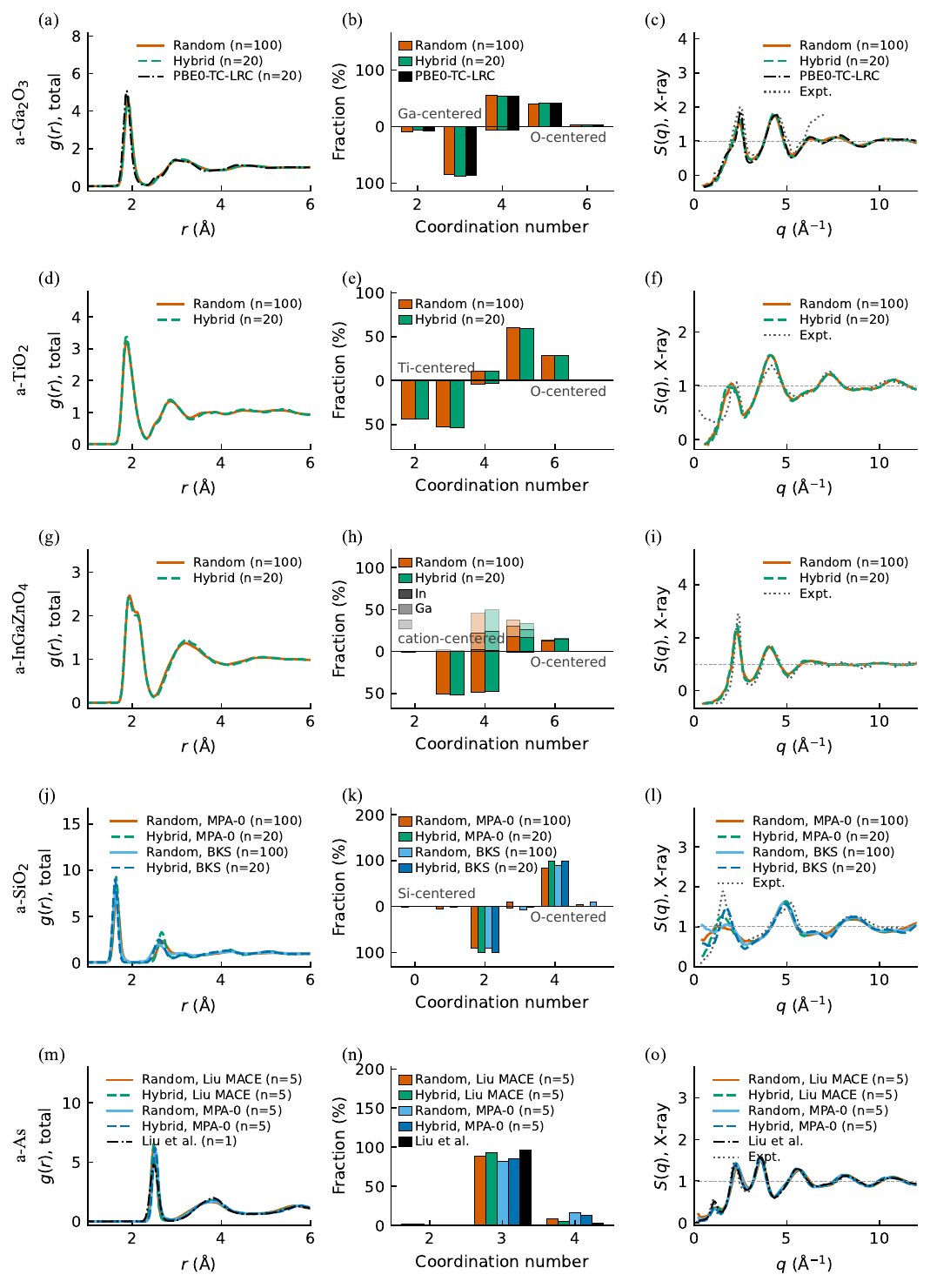}
\caption{Total $g(r)$ (left), coordination distributions (middle), and X-ray
$S(q)$ (right) of the random + relax (solid) and hybrid (dashed) ensembles:
(a--c) a-\ce{Ga2O3} with the PBE0-TC-LRC structures of
Ref.~\cite{kaewmeechai2025}; (d--f) a-\ce{TiO2}; (g--i) a-\ce{InGaZnO4};
(j--l) a-\ce{SiO2} with MACE-MPA-0 and BKS; (m--o) a-As with MACE-MPA-0,
the potential of Liu \textit{et al.}\ \cite{liu2026as}, and their
structure. Reference structures are dash-dotted. Cation- and O-centered
coordination fractions point up and down. $S(q)$ is averaged over 20
structures (five for a-As). Dotted: measured $S(q)$, from X-ray data for
a-\ce{Ga2O3} \cite{liu2023} and a-\ce{InGaZnO4} \cite{kumara2021} films and
for a-\ce{TiO2} \cite{petkov1998} and silica glass \cite{kohara2007} as
replotted in Refs.~\cite{yang2018tio2,zeidler2014}, and from combined
X-ray and neutron data for a-As \cite{smith1975} as replotted in
Ref.~\cite{liu2026as}; for one element, both give the same $S(q)$.}
\label{fig:grid}
\end{figure}

\begin{figure}[H]
\centering
\includegraphics[width=\textwidth]{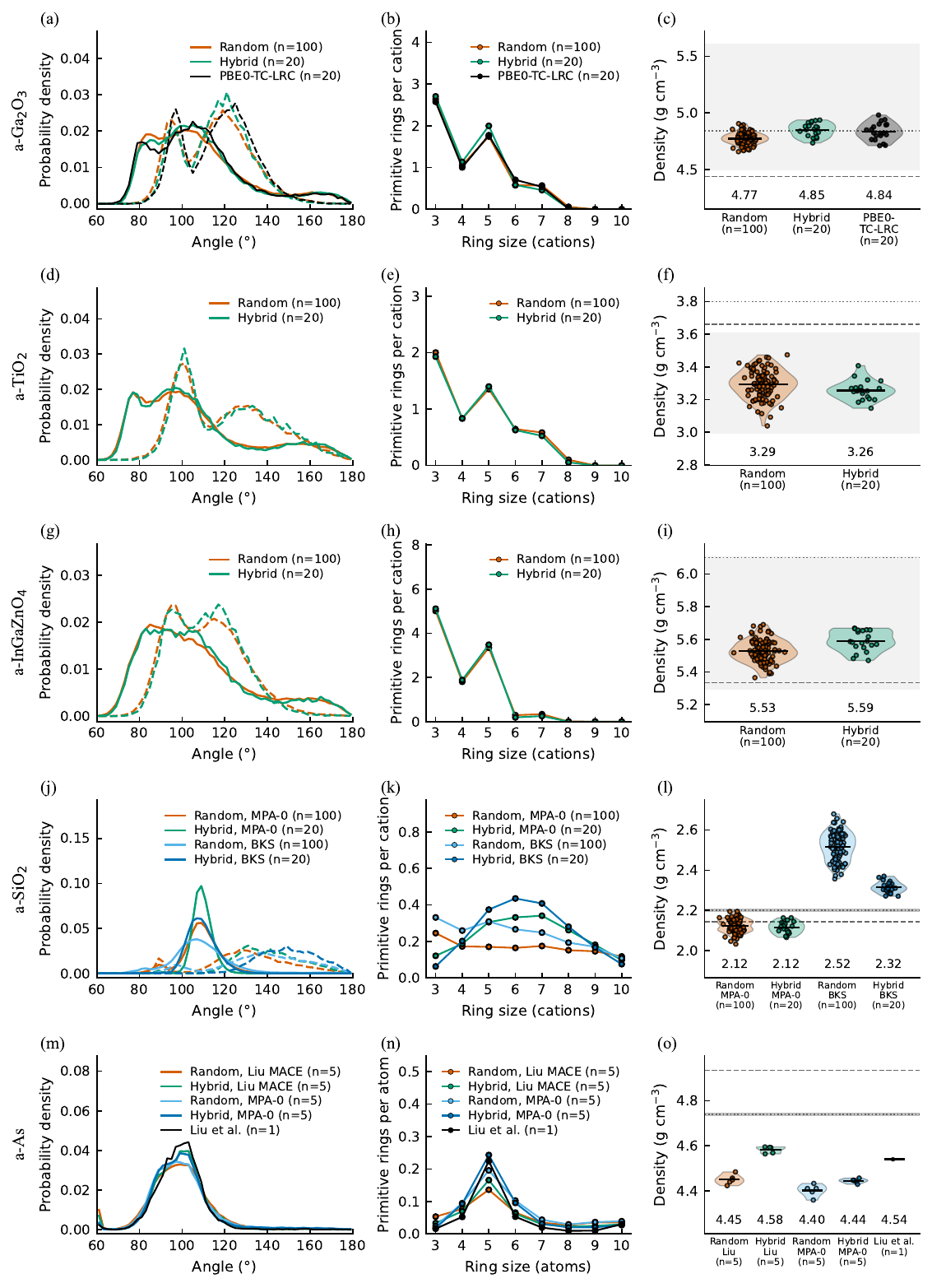}
\caption{Bond angles (left), primitive rings (middle), and density
(right) of the ensembles of Figure~\ref{fig:grid}. Angles: O--M--O
(solid) and M--O--M (dashed; As--As--As for a-As). Rings per cation (per
atom for a-As) \cite{goetzke1991,yuan2002}. In the metal oxides, they
mainly reflect the O coordination. Densities after cubic-cell relaxation, with the generation density
(dotted), $\rho_\mathrm{est}$ (dashed), and the measured value or
range (band).}
\label{fig:anneal}
\end{figure}

For the three metal oxides, random + relax gives cation coordination
within 0.5 of the references (Table~\ref{tab:oxides},
Figure~\ref{fig:grid}a--i). Both a-\ce{Ga2O3} ensembles have Ga--O
bonds \SI{0.03}{\angstrom} longer than the reference structures
\cite{kaewmeechai2025} and densities close to those of films sputtered
at room temperature, \SIrange{4.78}{4.84}{\gram\per\cubic\centi\metre}
\cite{han2020,liu2023}. For a-\ce{TiO2}, the Ti--O distance and
coordination agree with diffraction \cite{petkov1998}, and for
a-\ce{InGaZnO4}, the metal--O distances lie within
\SI{0.08}{\angstrom} of EXAFS \cite{kumara2021}.
Both a-\ce{TiO2} and a-\ce{InGaZnO4} relax into the measured film
density ranges, which vary with deposition conditions
\cite{mergel2000,ide2017,hu2021}, but below their densest films,
\SIrange{3.4}{3.6}{} and \SIrange{5.9}{6.1}{\gram\per\cubic\centi\metre}
(Figure~\ref{fig:anneal}c, f, i).

The X-ray $S(q)$ broadly follows the reference peak positions, but the
first peak is weaker for a-\ce{Ga2O3} and a-\ce{InGaZnO4}, and the weak,
broad first peak of a-\ce{TiO2} lies at slightly lower $q$. Relaxation also leaves O--O dimers, pairs of O atoms closer than
\SI{1.9}{\angstrom} (\suppref{sec:analysis}{Supplementary Section~S4}), in \SIlist{58;17;46}{\percent} of the a-\ce{Ga2O3},
a-\ce{TiO2}, and a-\ce{InGaZnO4} structures, and the hybrid route
removes all of them.
Detailed peak comparisons and reference uncertainties are given in
\suppref{sec:structural-comparisons}{Supplementary Section~S4.1}.

\begin{table}[H]
\centering
\caption{a-\ce{Ga2O3}, a-\ce{TiO2}, and a-\ce{InGaZnO4} ensembles with
MACE-MPA-0: random + relax (100 structures) and hybrid (20). Densities
(mean $\pm$ SD) are after cubic-cell relaxation from
\SIlist{4.84;3.80;6.10}{\gram\per\cubic\centi\metre}. CN, mean
coordination number.}
\label{tab:oxides}
\footnotesize
\resizebox{\textwidth}{!}{%
\begin{tabular}{@{}l l c c p{3.1cm}@{}}
\toprule
System & Descriptor & Random & Hybrid & Reference \\
\midrule
a-\ce{Ga2O3} & Ga--O (\si{\angstrom}) & 1.933 & 1.929 & 1.901 \cite{kaewmeechai2025} \\
 & CN(Ga); four-, five-fold (\si{\percent}) & 4.5; 55, 41 & 4.5; 55, 42 & 4.5; 55, 42 \cite{kaewmeechai2025} \\
 & O--Ga--O, Ga--O--Ga (\si{\degree}) & 107, 115 & 107, 116 & 107, 117 \cite{kaewmeechai2025} \\
 & density (\si{\gram\per\cubic\centi\metre}) & \num{4.77 \pm 0.05} & \num{4.85 \pm 0.06} & 4.5--5.6 \cite{passlack1995,nieminen1996} \\
\addlinespace
a-\ce{TiO2} & Ti--O (\si{\angstrom}) & 1.957 & 1.953 & 1.94--1.96 \cite{petkov1998} \\
 & CN(Ti) & 5.2 & 5.2 & 5.3 \cite{petkov1998} \\
 & O--Ti--O, Ti--O--Ti (\si{\degree}) & 105, 120 & 105, 121 & -- \\
 & density (\si{\gram\per\cubic\centi\metre}) & \num{3.29 \pm 0.09} & \num{3.26 \pm 0.06} & 3.0--3.6 \cite{mergel2000} \\
\addlinespace
a-\ce{InGaZnO4} & In--O, Ga--O, Zn--O (\si{\angstrom}) & 2.21, 1.93, 2.05 & 2.21, 1.92, 2.05 & 2.13, 1.87, 1.97 \cite{kumara2021} \\
 & CN(In), CN(Ga), CN(Zn) & 5.3, 4.4, 4.2 & 5.4, 4.3, 4.2 & 5.1, 3.9, 3.7 \cite{kumara2021} \\
 & density (\si{\gram\per\cubic\centi\metre}) & \num{5.53 \pm 0.07} & \num{5.59 \pm 0.06} & 5.3--6.1 \cite{ide2017,hu2021,kumara2021} \\
\bottomrule
\end{tabular}}
\end{table}

\newpage

\begin{table}[H]
\centering
\caption{Bond lengths, angles, coordination numbers, and
FSDP values of a-\ce{SiO2} and a-As. Densities (mean $\pm$ SD) are after a
separate cubic-cell relaxation. FSDP: position and height of the first
sharp diffraction peak (Figure~\ref{fig:grid}l, o). The a-As reference
values are computed from the deposited model structure.}
\label{tab:glasses}
\footnotesize
\resizebox{\textwidth}{!}{%
\begin{tabular}{@{}l c c c c c@{}}
\toprule
a-\ce{SiO2} & \multicolumn{2}{c}{MACE-MPA-0} & \multicolumn{2}{c}{BKS} & Reference \\
\cmidrule(lr){2-3} \cmidrule(lr){4-5}
 & Random & Hybrid & Random & Hybrid & Expt.\ \cite{mozzi1969,wright1994,kohara2007} \\
\midrule
Si--O (\si{\angstrom}) & 1.657 & 1.636 & 1.635 & 1.615 & 1.608--1.62 \\
CN(Si); fourfold (\si{\percent}) & 3.9; 83 & 4.0; 99.3 & 4.1; 89 & 4.0; 99 & 4; $\sim$100 \\
Si--O--Si (\si{\degree}) & 130 & 140 & 138 & 149 & 144--152 \\
density (\si{\gram\per\cubic\centi\metre}) & \num{2.12 \pm 0.03} & \num{2.12 \pm 0.03} & \num{2.52 \pm 0.07} & \num{2.32 \pm 0.03} & 2.20 \\
FSDP position (\si{\per\angstrom}) / height & none & 1.50 / 1.26 & 1.70 / 1.06 & 1.74 / 1.48 & 1.55 / 1.9 \\
\midrule
a-As & \multicolumn{2}{c}{MACE of Liu \textit{et al.}} & \multicolumn{2}{c}{MACE-MPA-0} & Reference \\
\cmidrule(lr){2-3} \cmidrule(lr){4-5}
 & Random & Hybrid & Random & Hybrid & Model \cite{liu2026as} \\
\midrule
As--As (\si{\angstrom}) & 2.496 & 2.493 & 2.538 & 2.533 & 2.498 \\
CN(As); threefold (\si{\percent}) & 3.1; 89.3 & 3.0; 93.1 & 3.2; 82.6 & 3.1; 85.5 & 3.0; 96.4 \\
As--As--As (\si{\degree}) & 98.1 & 98.3 & 99.5 & 99.4 & 98.6 \\
density (\si{\gram\per\cubic\centi\metre}) & \num{4.45 \pm 0.02} & \num{4.58 \pm 0.02} & \num{4.40 \pm 0.03} & \num{4.44 \pm 0.01} & 4.54 \\
FSDP position (\si{\per\angstrom}) / height & 1.18 / 0.37 & 1.18 / 0.40 & 1.10 / 0.33 & 1.10 / 0.37 & 1.10 / 0.51 \\
\bottomrule
\end{tabular}}
\end{table}
Neighbor cutoffs, ring definitions, and structure-factor processing
are given in \suppref{sec:analysis}{Supplementary Section~S4}.
With MACE-MPA-0, the relaxed and annealed a-\ce{SiO2} ensembles have
similar radial distribution functions, but relaxation alone leaves
coordination defects and no first sharp diffraction peak (FSDP)
(Table~\ref{tab:glasses}, Figure~\ref{fig:grid}j--l).
Annealing raises the four-fold Si fraction to \SI{99.3}{\percent},
leaves 16 of the 20 structures free of O--O dimers, which every relaxed
structure contains, and removes the low-$q$ rise of $S(q)$. It also changes the rings
(Figure~\ref{fig:anneal}k): relaxation alone leaves a flat distribution
with many three-membered rings, whereas after the anneal the
distribution peaks at six- to seven-membered rings, most clearly with
BKS, close to simulations of silica glass, in which six-membered rings are the most frequent \cite{vollmayr1996}.
With MACE-MPA-0, bond lengths and angles remain outside the diffraction
ranges \cite{mozzi1969,wright1994}, and the FSDP is weaker than measured
\cite{kohara2007}. Starting from the same placements,
BKS gives local descriptors within those ranges after annealing,
but its FSDP remains weak and shifted. The comparison involves both
potential and density: BKS was annealed at
\SI{2.20}{\gram\per\cubic\centi\metre}, MACE-MPA-0 at
\SI{2.12}{\gram\per\cubic\centi\metre}
(\suppref{sec:provenance}{Supplementary Section~S3}).

For a-As, the system-specific MACE potential of Liu \textit{et al.}\
\cite{liu2026as} reproduces their model's bond length and angle after
relaxation alone (Table~\ref{tab:glasses}, Figure~\ref{fig:grid}m--o).
Annealing reduces excess four-fold sites, increases five-membered
rings (Figure~\ref{fig:anneal}n), and removes the low-$q$ rise of $S(q)$, whereas MACE-MPA-0 retains
more coordination defects. All four ensembles remain less dense
than experiment, and even the system-specific potential gives an
annealed FSDP weaker than both the published model and experiment
\cite{smith1975}. Our values average five 2000-atom cells; Liu
\textit{et al.}\ deposited one 2000-atom model and estimated
uncertainties from ten independent runs of that size \cite{liu2026as}.
Including the quench and low-temperature equilibration, our hybrid
protocol lasts \SI{72}{\pico\second}, about 20 times shorter than
theirs, and still recovers the local structure. Preparation settings and further ring and peak
comparisons are given in
\suppref{sec:provenance}{Supplementary Section~S3} and
\suppref{sec:structural-comparisons}{Section~S4.1}.

\begin{figure}[t]
\centering
\includegraphics[width=0.98\textwidth]{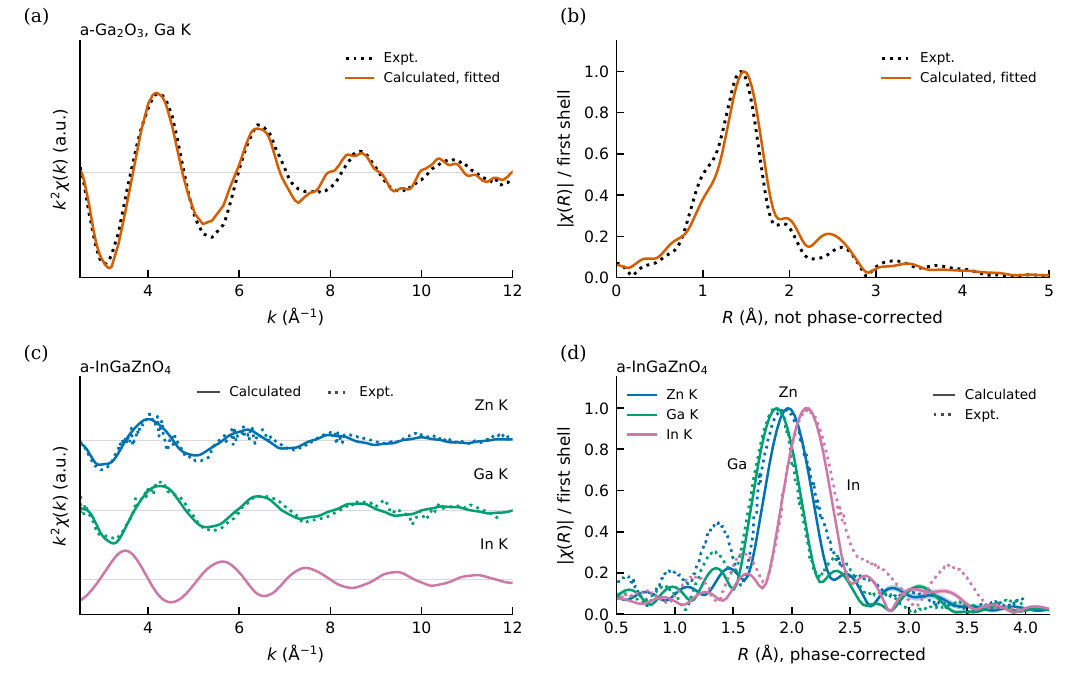}
\caption{K-edge EXAFS calculated with FEFF8.5L from static MACE-MPA-0
structures (solid) against measured spectra (dotted). (a) Ga K-edge
$k^2\chi(k)$ of 100 a-\ce{Ga2O3} random + relax cells, with a scale, an
offset, an energy shift, and an extra damping $\Delta\sigma^2$ fitted to
the as-deposited film of Reis \textit{et al.}\ \cite{reis2023}.
(b) Fourier transforms of (a), not phase-corrected. (c, d) Zn, Ga, and
In K edges of 504 a-\ce{InGaZnO4} cells relaxed at
\SI{6.10}{\gram\per\cubic\centi\metre}. (c) $k^2\chi(k)$, offset for
clarity. Zn and Ga are fitted as in (a) to a film of different
composition (In:Ga:Zn = 3:0.4:2) \cite{revenant2015}, so (c) tests only
the first-shell phase and frequency. In is shown unfitted, as no
measured spectrum is published. (d) Phase-corrected $|\chi(R)|$ of the
$k^3$-weighted spectra against Kumara \textit{et al.}\ \cite{kumara2021},
whose $k$ range is not stated, scaled to the first-shell maximum.
Windows are given in Supplementary Section~S6.1.}
\label{fig:exafs}
\end{figure}

EXAFS spectra calculated with FEFF8.5L (Feff8L) \cite{feff85exafs} and averaged
over all metal atoms allow direct comparison with X-ray absorption
measurements (Figure~\ref{fig:exafs}). The spectra are calculated from random + relax
structures, so thermal vibrations are not included. The Fourier transforms are converged to within \SI{2}{\percent} up to
\SI{4}{\angstrom}, while longer scattering paths still change the raw
$\chi(k)$ slightly
(\suppref{sec:exafs-convergence}{Supplementary Section~S6.4}).

For a-\ce{Ga2O3}, the spectrum is averaged over 100 random + relax
cells made as in Table~\ref{tab:oxides}
(\suppref{sec:exafs-ga2o3}{Supplementary Section~S6.3}). After fitting a scale, an offset, an energy shift, and an extra damping term $\Delta\sigma^2$, it matches the measured spectrum \cite{reis2023} with a
normalized squared residual of 0.043 over
$k = \SIrange{2.5}{12}{\per\angstrom}$ and reproduces the dominant
Ga--O shell.

For a-\ce{InGaZnO4}, the spectra are averaged over 504 cells made as in
Table~\ref{tab:oxides} but relaxed at fixed volume, at the measured
density of \SI{6.10}{\gram\per\cubic\centi\metre}. The Zn and Ga
spectra are fitted in the same way to a film of different composition
\cite{revenant2015}, and no measured In spectrum is available. Against
the phase-corrected transforms of Kumara \textit{et al.}\ \cite{kumara2021}, the
first-shell maxima agree within \SI{0.04}{\angstrom} (the longer mean distances of Table~\ref{tab:oxides} are expected for asymmetric shells), but the In feature near
\SI{3.3}{\angstrom} is too weak, even after annealing
(\suppref{sec:exafs-igzo}{Supplementary Sections~S6.5}--\suppref{sec:exafs-annealed}{S6.7}).

These tests suggest how to choose a route. Random + relax is the
quickest (\suppref{tab:timings}{Supplementary Table~S5}). For the
metal oxides, it already gives the cation
coordination.
It does, however, leave O--O dimers in \SIrange{17}{58}{\percent} of these structures.
The hybrid route removes them and is the better choice when high-quality
structures are needed, for example, as training data. For a-\ce{SiO2}
and a-As, the anneal is required because relaxation alone leaves
\SI{17}{\percent} of Si and \SIrange{11}{17}{\percent} of As wrongly coordinated. Even after the
anneal, their medium-range order stays below experimental values with this
protocol. For EXAFS, the number of cells needed for a converged mean
differs between systems, from tens for a-\ce{Ga2O3} to hundreds for
a-\ce{InGaZnO4}, at about 0.1 core-hour of FEFF per cell, so the
cell-sampling check of \suppref{sec:exafs-convergence}{Supplementary Section~S6.4} should be run for each
new system.

Foundation potentials need no training and cover most of the periodic
table, so a new composition can be modeled at once. In our tests,
however, MACE-MPA-0 gave slightly longer metal--O bonds than the
references and oxide densities below those of the densest films. After relaxation, it also left fewer four-fold Si than BKS in
a-\ce{SiO2} and fewer three-fold As than the potential of
Liu \textit{et al.}\ in a-As. 
We therefore recommend checking a few structures with DFT or using a
fine-tuned or system-specific potential, which AmorphGen accepts
without any change to the workflow
\cite{tompa2026finetuning,fragapane2026amorphous}.
The melt-quench route was not benchmarked separately. Its stages 4--7
run the same code as the hybrid route, which was tested above at fixed
volume; stages 1--3, which optimize, equilibrate, and heat the starting
crystal, were not tested. Its example settings are in
\suppref{sec:provenance}{Supplementary Section~S3}. With foundation
potentials, it should be run at fixed volume, with the barostat applied
only at low temperature \cite{li2026meltquench}.

\section{Impact}
\label{impact}

AmorphGen makes ensemble studies of amorphous materials practical by
combining structure generation, relaxation, and analysis in one
workflow. In Section~\ref{examples}, generating and relaxing 100
structures of 270 to 648 atoms with MACE-MPA-0 took between
\SI{48}{\minute} and \SI{3}{\hour} per oxide on one NVIDIA A100 GPU,
including placement (\suppref{tab:timings}{Supplementary Table~S5}).
Many compositions can therefore be screened before more expensive
annealing or DFT calculations.

Potentials trained for a specific material can also be used within the
same workflow. With the MACE potential of Liu \textit{et al.}\
\cite{liu2026as}, trained on a-As, five 2000-atom structures were
annealed and quenched in about \SI{10}{\hour} on one GPU and compared
directly with their model (Table~\ref{tab:glasses}).

The FEFF interface compares these ensembles directly with X-ray
absorption measurements, element by element. For a-\ce{Ga2O3} and
a-\ce{InGaZnO4}, the calculated spectra reproduce the measured first
shells of each metal, and they also show where a model differs beyond
the first shell (Figure~\ref{fig:exafs}).

\section{Conclusions}
\label{conclusions}

We developed AmorphGen, a Python package that automates amorphous
structure generation. Given a chemical composition, it places atoms
randomly under constraints derived from that composition, then relaxes
the structures or anneals them further with the hybrid route. It also
supports the full melt-quench pipeline starting from a crystal.
All routes run with machine-learning or classical interatomic
potentials.

AmorphGen also includes an analysis module, which computes radial
distribution functions, coordination, bond angles, rings, polyhedral
connectivity, X-ray or neutron structure factors, and, with an external
FEFF executable, ensemble-averaged EXAFS spectra. It also compares each
set of structures with reference data.

We validated the generated structures on three metal oxides, on
silica glass (\ce{SiO2}), and on As. For these three oxides, random + relax alone gives cation coordination
close to the references. \ce{SiO2} and As need the anneal, which
reduces their coordination defects, but the medium-range order stays
below experiment with this protocol. The EXAFS spectra of a-\ce{Ga2O3} and a-\ce{InGaZnO4}
capture the dominant measured first-shell features.

\section*{Acknowledgements}
\label{acknowledgements}
C.K. and D.O.S. acknowledge the PRAETORIAN project, funded by UK Research and Innovation (UKRI) under the UK government's Horizon Europe funding guarantee (EP/Y019504/1). 
The computations described in this research were performed using the
Baskerville Tier 2 HPC service (\url{https://www.baskerville.ac.uk/}). Baskerville was funded by the
EPSRC and UKRI through the World Class Labs scheme (EP/T022221/1) and the
Digital Research Infrastructure programme (EP/W032244/1) and is operated
by Advanced Research Computing at the University of Birmingham. The computations described
in this paper were also performed using the University of Birmingham's
BlueBEAR HPC service, which provides a high-performance computing
service to the University's research community (see
\url{http://www.birmingham.ac.uk/bear} for more details), and the Sulis
Tier 2 HPC platform hosted by the Scientific Computing Research
Technology Platform at the University of Warwick (funded by EPSRC Grant
EP/T022108/1 and the HPC Midlands+ consortium). Through our membership
of the UK's HEC Materials Chemistry Consortium, which is funded by the
UK Engineering and Physical Sciences Research Council (EPSRC;
EP/L000202, EP/R029431, EP/T022213, EP/X035859), this work also used the ARCHER2 UK
National Supercomputing Service. We are also grateful to the UK
Materials and Molecular Modelling Hub for computational resources, which
is partially funded by EPSRC (EP/T022213/1, EP/W032260/1 and
EP/P020194/1). The structures
in Figure~\ref{fig:overview} were rendered with hofmann
\cite{morganHofmann}.

\section*{Data availability}

The software is archived on Zenodo as AmorphGen v1.0.0
(\url{https://doi.org/10.5281/zenodo.24207927}). The data behind
Section~\ref{examples} are archived at
\url{https://doi.org/10.5281/zenodo.24284240}.

\section*{Supplementary material}

The supplementary material details the composition rules and material
classes (\suppref{sec:composition}{Section~S1},
\suppref{tab:packing}{Table~S1}); the density calibration and its
reference data (\suppref{sec:packing-calibration}{Section~S2},
Tables~\suppref{tab:reference-densities}{S2}--\suppref{tab:packing-fit}{S3}
and \suppref{fig:parity}{Figure~S1}); the structural ensembles,
precision, seeds, and task splits (\suppref{sec:provenance}{Section~S3},
\suppref{tab:ensemble-provenance}{Table~S4}); the structural analysis
settings (\suppref{sec:analysis}{Section~S4}); the computational cost
of structure generation (\suppref{sec:cost}{Section~S5},
\suppref{tab:timings}{Table~S5}); and the EXAFS protocol, validation,
convergence tests, and comparisons with experiment
(\suppref{sec:exafs}{Section~S6},
Tables~\suppref{tab:exafs-convergence}{S6}--\suppref{tab:exafs-igzo-features}{S8}).

\section*{CRediT authorship contribution statement}

\textbf{Chaiyawat Kaewmeechai:} Conceptualization, Data curation,
Formal analysis, Investigation, Methodology, Software, Validation,
Visualization, Writing -- original draft, Writing -- review \& editing.
\textbf{Louie Slocombe:} Software, Validation,
Project administration,
Writing -- review \& editing.
\textbf{Alexis G. Manche:} Formal analysis, Investigation, Methodology,
Software, Validation, Writing -- review \& editing.
\textbf{David O. Scanlon:} Conceptualization, Funding acquisition,
Project administration, Resources, Supervision,
Writing -- review \& editing.

\section*{Declaration of competing interest}

The authors declare that they have no known competing financial
interests or personal relationships that could have appeared to
influence the work reported in this paper.

\section*{Declaration of generative AI and AI-assisted technologies in the writing process}

During the preparation of this work, the authors used Claude Code
with Claude (Anthropic) to draft and revise text and to check the software
descriptions against the AmorphGen source code, and Grammarly to check
spelling, grammar, and style. After using these tools, the authors
reviewed and edited the content as needed and take full responsibility
for the content of the published article.

\bibliographystyle{elsarticle-num}
\bibliography{refs}

\begin{thebibliography}{10}
\expandafter\ifx\csname url\endcsname\relax
  \def\url#1{\texttt{#1}}\fi
\expandafter\ifx\csname urlprefix\endcsname\relax\def\urlprefix{URL }\fi
\expandafter\ifx\csname href\endcsname\relax
  \def\href#1#2{#2} \def\path#1{#1}\fi

\bibitem{pickard2011}
C.~J. Pickard, R.~J. Needs, Ab initio random structure searching, Journal of Physics: Condensed Matter 23 (2011) 053201.
\newblock \href {https://doi.org/10.1088/0953-8984/23/5/053201} {\path{doi:10.1088/0953-8984/23/5/053201}}.

\bibitem{youn2014}
Y.~Youn, Y.~Kang, S.~Han, An efficient method to generate amorphous structures based on local geometry, Computational Materials Science 95 (2014) 256--262.
\newblock \href {https://doi.org/10.1016/j.commatsci.2014.07.053} {\path{doi:10.1016/j.commatsci.2014.07.053}}.

\bibitem{jones2020glassy}
E.~B. Jones, V.~Stevanovi{\'c}, The glassy solid as a statistical ensemble of crystalline microstates, npj Computational Materials 6 (2020) 56.
\newblock \href {https://doi.org/10.1038/s41524-020-0329-2} {\path{doi:10.1038/s41524-020-0329-2}}.

\bibitem{stich1991amorphous}
I.~{\v{S}}tich, R.~Car, M.~Parrinello, Amorphous silicon studied by ab initio molecular dynamics: Preparation, structure, and properties, Physical Review B 44 (1991) 11092--11104.
\newblock \href {https://doi.org/10.1103/PhysRevB.44.11092} {\path{doi:10.1103/PhysRevB.44.11092}}.

\bibitem{martinez2009packmol}
L.~Mart{\'i}nez, R.~Andrade, E.~G. Birgin, J.~M. Mart{\'i}nez, {PACKMOL}: a package for building initial configurations for molecular dynamics simulations, Journal of Computational Chemistry 30 (2009) 2157--2164.
\newblock \href {https://doi.org/10.1002/jcc.21224} {\path{doi:10.1002/jcc.21224}}.

\bibitem{wooten1985}
F.~Wooten, K.~Winer, D.~Weaire, Computer generation of structural models of amorphous {Si} and {Ge}, Physical Review Letters 54 (1985) 1392--1395.
\newblock \href {https://doi.org/10.1103/PhysRevLett.54.1392} {\path{doi:10.1103/PhysRevLett.54.1392}}.

\bibitem{barkema2000crn}
G.~T. Barkema, N.~Mousseau, High-quality continuous random networks, Physical Review B 62 (2000) 4985--4990.
\newblock \href {https://doi.org/10.1103/PhysRevB.62.4985} {\path{doi:10.1103/PhysRevB.62.4985}}.

\bibitem{zheng2025gbsmc}
Z.~Zheng, F.~Zheng, Y.-N. Wu, S.~Chen, L.-W. Wang, A generalized bond switching {Monte Carlo} method for amorphous structure generation, Computational Materials Today 6 (2025) 100031.
\newblock \href {https://doi.org/10.1016/j.commt.2025.100031} {\path{doi:10.1016/j.commt.2025.100031}}.

\bibitem{mcgreevy1988rmc}
R.~L. McGreevy, L.~Pusztai, Reverse {Monte Carlo} simulation: a new technique for the determination of disordered structures, Molecular Simulation 1 (1988) 359--367.
\newblock \href {https://doi.org/10.1080/08927028808080958} {\path{doi:10.1080/08927028808080958}}.

\bibitem{tucker2007rmcprofile}
M.~G. Tucker, D.~A. Keen, M.~T. Dove, A.~L. Goodwin, Q.~Hui, {RMCProfile}: reverse {Monte Carlo} for polycrystalline materials, Journal of Physics: Condensed Matter 19 (2007) 335218.
\newblock \href {https://doi.org/10.1088/0953-8984/19/33/335218} {\path{doi:10.1088/0953-8984/19/33/335218}}.

\bibitem{aoun2016fullrmc}
B.~Aoun, {Fullrmc}, a rigid body reverse {Monte Carlo} modeling package enabled with machine learning and artificial intelligence, Journal of Computational Chemistry 37 (2016) 1102--1111.
\newblock \href {https://doi.org/10.1002/jcc.24304} {\path{doi:10.1002/jcc.24304}}.

\bibitem{opletal2008hrmc}
G.~Opletal, T.~C. Petersen, B.~O'Malley, I.~K. Snook, D.~G. McCulloch, I.~Yarovsky, {HRMC}: hybrid reverse {Monte Carlo} method with silicon and carbon potentials, Computer Physics Communications 178 (2008) 777--787.
\newblock \href {https://doi.org/10.1016/j.cpc.2007.12.007} {\path{doi:10.1016/j.cpc.2007.12.007}}.

\bibitem{gore2026torchdisorder}
A.~Gore, X.~Gouws, C.~G. {Tetsassi Feugmo}, {TorchDisorder}: a differentiable framework for generating physically realistic disorder structures from experimental diffraction data, Journal of Chemical Theory and Computation 22 (2026) 6859--6873.
\newblock \href {https://doi.org/10.1021/acs.jctc.6c00490} {\path{doi:10.1021/acs.jctc.6c00490}}.

\bibitem{thompson2022lammps}
A.~P. Thompson, H.~M. Aktulga, R.~Berger, D.~S. Bolintineanu, W.~M. Brown, P.~S. Crozier, P.~J. {in 't Veld}, A.~Kohlmeyer, S.~G. Moore, T.~D. Nguyen, et~al., {LAMMPS} -- a flexible simulation tool for particle-based materials modeling at the atomic, meso, and continuum scales, Computer Physics Communications 271 (2022) 108171.
\newblock \href {https://doi.org/10.1016/j.cpc.2021.108171} {\path{doi:10.1016/j.cpc.2021.108171}}.

\bibitem{hirel2015atomsk}
P.~Hirel, {Atomsk}: a tool for manipulating and converting atomic data files, Computer Physics Communications 197 (2015) 212--219.
\newblock \href {https://doi.org/10.1016/j.cpc.2015.07.012} {\path{doi:10.1016/j.cpc.2015.07.012}}.

\bibitem{janssen2019}
J.~Janssen, S.~Surendralal, Y.~Lysogorskiy, M.~Todorova, T.~Hickel, R.~Drautz, J.~Neugebauer, {pyiron}: An integrated development environment for computational materials science, Computational Materials Science 163 (2019) 24--36.
\newblock \href {https://doi.org/10.1016/j.commatsci.2018.07.043} {\path{doi:10.1016/j.commatsci.2018.07.043}}.

\bibitem{cottom2026vitriflow}
J.~Cottom, R.~Delhomme, E.~Olsson, {Vitriflow}: Calibrated amorphous structure ensembles from melt--quench simulations, Computational Materials Science 275 (2026) 115098.
\newblock \href {https://doi.org/10.1016/j.commatsci.2026.115098} {\path{doi:10.1016/j.commatsci.2026.115098}}.

\bibitem{amorphouspy}
L.~Talirz, A.~Atila, M.~Sadowski, J.~Janssen, \href{https://github.com/glasagent/amorphouspy}{{amorphouspy}: workflows for computing glass properties at the atomic level}, version 0.8.1 (2026).
\newblock \href {https://doi.org/10.5281/zenodo.21365089} {\path{doi:10.5281/zenodo.21365089}}.
\newline\urlprefix\url{https://github.com/glasagent/amorphouspy}

\bibitem{aykol2018mpmorph}
M.~Aykol, S.~S. Dwaraknath, W.~Sun, K.~A. Persson, Thermodynamic limit for synthesis of metastable inorganic materials, Science Advances 4 (2018) eaaq0148.
\newblock \href {https://doi.org/10.1126/sciadv.aaq0148} {\path{doi:10.1126/sciadv.aaq0148}}.

\bibitem{vanbeest1990bks}
B.~W.~H. van Beest, G.~J. Kramer, R.~A. van Santen, Force fields for silicas and aluminophosphates based on ab initio calculations, Physical Review Letters 64 (1990) 1955--1958.
\newblock \href {https://doi.org/10.1103/PhysRevLett.64.1955} {\path{doi:10.1103/PhysRevLett.64.1955}}.

\bibitem{carre2008chik}
A.~Carr{\'e}, J.~Horbach, S.~Ispas, W.~Kob, New fitting scheme to obtain effective potential from {Car-Parrinello} molecular-dynamics simulations: application to silica, EPL (Europhysics Letters) 82 (2008) 17001.
\newblock \href {https://doi.org/10.1209/0295-5075/82/17001} {\path{doi:10.1209/0295-5075/82/17001}}.

\bibitem{oeffner1998}
R.~D. Oeffner, S.~R. Elliott, Interatomic potential for germanium dioxide empirically fitted to an ab initio energy surface, Physical Review B 58 (1998) 14791--14803.
\newblock \href {https://doi.org/10.1103/PhysRevB.58.14791} {\path{doi:10.1103/PhysRevB.58.14791}}.

\bibitem{batatia2022mace}
I.~Batatia, D.~P. Kov{\'a}cs, G.~N.~C. Simm, C.~Ortner, G.~Cs{\'a}nyi, {MACE}: Higher order equivariant message passing neural networks for fast and accurate force fields, in: Advances in Neural Information Processing Systems, Vol.~35, 2022, pp. 11423--11436.
\newblock \href {https://doi.org/10.52202/068431-0830} {\path{doi:10.52202/068431-0830}}.

\bibitem{batatia2023foundation}
I.~Batatia, P.~Benner, Y.~Chiang, A.~M. Elena, D.~P. Kov{\'a}cs, J.~Riebesell, X.~R. Advincula, M.~Asta, M.~Avaylon, W.~J. Baldwin, et~al., A foundation model for atomistic materials chemistry, The Journal of Chemical Physics 163 (2025) 184110.
\newblock \href {https://doi.org/10.1063/5.0297006} {\path{doi:10.1063/5.0297006}}.

\bibitem{deng2023chgnet}
B.~Deng, P.~Zhong, K.~Jun, J.~Riebesell, K.~Han, C.~J. Bartel, G.~Ceder, {CHGNet} as a pretrained universal neural network potential for charge-informed atomistic modelling, Nature Machine Intelligence 5 (2023) 1031--1041.
\newblock \href {https://doi.org/10.1038/s42256-023-00716-3} {\path{doi:10.1038/s42256-023-00716-3}}.

\bibitem{park2024sevennet}
Y.~Park, J.~Kim, S.~Hwang, S.~Han, Scalable parallel algorithm for graph neural network interatomic potentials in molecular dynamics simulations, Journal of Chemical Theory and Computation 20 (2024) 4857--4868.
\newblock \href {https://doi.org/10.1021/acs.jctc.4c00190} {\path{doi:10.1021/acs.jctc.4c00190}}.

\bibitem{li2025finitet}
D.~Li, J.~Yang, X.~Chen, L.~Yu, S.~Liu, Are foundational atomistic models reliable for finite-temperature molecular dynamics?, The Journal of Physical Chemistry C 129 (2025) 21538--21544.
\newblock \href {https://doi.org/10.1021/acs.jpcc.5c07541} {\path{doi:10.1021/acs.jpcc.5c07541}}.

\bibitem{li2026meltquench}
S.~Li, Y.~An, X.~Guo, W.~Yang, Z.~Wang, Melt-quench failures and practical solutions for universal machine-learning interatomic potentials in amorphous structure generation, preprint (2026).
\newblock \href {http://arxiv.org/abs/2606.16385} {\path{arXiv:2606.16385}}, \href {https://doi.org/10.48550/arXiv.2606.16385} {\path{doi:10.48550/arXiv.2606.16385}}.

\bibitem{fragapane2026amorphous}
N.~L. Fragapane, V.~L. Deringer, Amorphous materials as a frontier challenge for universal interatomic potentials, preprint (2026).
\newblock \href {http://arxiv.org/abs/2607.11384} {\path{arXiv:2607.11384}}, \href {https://doi.org/10.48550/arXiv.2607.11384} {\path{doi:10.48550/arXiv.2607.11384}}.

\bibitem{tompa2026finetuning}
T.~L. Tompa, E.~Varga-Umbrich, I.~Batatia, A.~M. Elena, N.~Bernstein, G.~Cs{\'a}nyi, Fine-tuning {MLIP} foundation models: strategies for accuracy and transferability, preprint (2026).
\newblock \href {http://arxiv.org/abs/2606.12704} {\path{arXiv:2606.12704}}, \href {https://doi.org/10.48550/arXiv.2606.12704} {\path{doi:10.48550/arXiv.2606.12704}}.

\bibitem{deringer2018gap}
V.~L. Deringer, G.~Cs{\'a}nyi, Machine learning based interatomic potential for amorphous carbon, Physical Review B 95 (2017) 094203.
\newblock \href {https://doi.org/10.1103/PhysRevB.95.094203} {\path{doi:10.1103/PhysRevB.95.094203}}.

\bibitem{deringer2019adma}
V.~L. Deringer, M.~A. Caro, G.~Cs{\'a}nyi, Machine learning interatomic potentials as emerging tools for materials science, Advanced Materials 31 (2019) 1902765.
\newblock \href {https://doi.org/10.1002/adma.201902765} {\path{doi:10.1002/adma.201902765}}.

\bibitem{liu2025autoplex}
Y.~Liu, J.~D. Morrow, C.~Ertural, N.~L. Fragapane, J.~L.~A. Gardner, A.~A. Naik, Y.~Zhou, J.~George, V.~L. Deringer, An automated framework for exploring and learning potential-energy surfaces, Nature Communications 16 (2025) 7666.
\newblock \href {https://doi.org/10.1038/s41467-025-62510-6} {\path{doi:10.1038/s41467-025-62510-6}}.

\bibitem{leroux2010rings}
S.~Le~Roux, P.~Jund, Ring statistics analysis of topological networks: New approach and application to amorphous {GeS$_2$} and {SiO$_2$} systems, Computational Materials Science 49 (2010) 70--83.
\newblock \href {https://doi.org/10.1016/j.commatsci.2010.04.023} {\path{doi:10.1016/j.commatsci.2010.04.023}}.

\bibitem{stukowski2010ovito}
A.~Stukowski, Visualization and analysis of atomistic simulation data with {OVITO}--the open visualization tool, Modelling and Simulation in Materials Science and Engineering 18 (2010) 015012.
\newblock \href {https://doi.org/10.1088/0965-0393/18/1/015012} {\path{doi:10.1088/0965-0393/18/1/015012}}.

\bibitem{larsen2017}
A.~{Hjorth Larsen}, J.~J. Mortensen, J.~Blomqvist, I.~E. Castelli, R.~Christensen, M.~Du{\l}ak, J.~Friis, M.~N. Groves, B.~Hammer, C.~Hargus, et~al., The atomic simulation environment --- a {Python} library for working with atoms, Journal of Physics: Condensed Matter 29 (2017) 273002.
\newblock \href {https://doi.org/10.1088/1361-648X/aa680e} {\path{doi:10.1088/1361-648X/aa680e}}.

\bibitem{ankudinov1998feff8}
A.~L. Ankudinov, B.~Ravel, J.~J. Rehr, S.~D. Conradson, Real-space multiple-scattering calculation and interpretation of {X}-ray-absorption near-edge structure, Physical Review B 58 (1998) 7565--7576.
\newblock \href {https://doi.org/10.1103/PhysRevB.58.7565} {\path{doi:10.1103/PhysRevB.58.7565}}.

\bibitem{rehr2000rmp}
J.~J. Rehr, R.~C. Albers, Theoretical approaches to {X}-ray absorption fine structure, Reviews of Modern Physics 72 (2000) 621--654.
\newblock \href {https://doi.org/10.1103/RevModPhys.72.621} {\path{doi:10.1103/RevModPhys.72.621}}.

\bibitem{reis2023}
A.~Reis, M.~Hanke, O.~Bierwagen, A.~Trampert, P.~Mazzolini, E.~Welter, Disorder--order transition in {Ga$_2$O$_3$} and its solid solution with {In$_2$O$_3$} upon thermal annealing, physica status solidi (b) 260 (2023) 2200535.
\newblock \href {https://doi.org/10.1002/pssb.202200535} {\path{doi:10.1002/pssb.202200535}}.

\bibitem{harris2020numpy}
C.~R. Harris, K.~J. Millman, S.~J. van~der Walt, R.~Gommers, P.~Virtanen, D.~Cournapeau, E.~Wieser, J.~Taylor, S.~Berg, N.~J. Smith, et~al., Array programming with {NumPy}, Nature 585 (2020) 357--362.
\newblock \href {https://doi.org/10.1038/s41586-020-2649-2} {\path{doi:10.1038/s41586-020-2649-2}}.

\bibitem{virtanen2020scipy}
P.~Virtanen, R.~Gommers, T.~E. Oliphant, M.~Haberland, T.~Reddy, D.~Cournapeau, E.~Burovski, P.~Peterson, W.~Weckesser, J.~Bright, et~al., {SciPy} 1.0: fundamental algorithms for scientific computing in {Python}, Nature Methods 17 (2020) 261--272.
\newblock \href {https://doi.org/10.1038/s41592-019-0686-2} {\path{doi:10.1038/s41592-019-0686-2}}.

\bibitem{pyyaml}
{PyYAML developers}, \href{https://github.com/yaml/pyyaml}{{PyYAML}: a {YAML} parser and emitter for {Python}}, accessed 13 November 2026 (2026).
\newline\urlprefix\url{https://github.com/yaml/pyyaml}

\bibitem{hunter2007}
J.~D. Hunter, {Matplotlib}: A {2D} graphics environment, Computing in Science \& Engineering 9 (2007) 90--95.
\newblock \href {https://doi.org/10.1109/MCSE.2007.55} {\path{doi:10.1109/MCSE.2007.55}}.

\bibitem{paszke2019pytorch}
A.~Paszke, S.~Gross, F.~Massa, A.~Lerer, J.~Bradbury, G.~Chanan, T.~Killeen, Z.~Lin, N.~Gimelshein, L.~Antiga, et~al., \href{https://proceedings.neurips.cc/paper/2019/hash/bdbca288fee7f92f2bfa9f7012727740-Abstract.html}{{PyTorch}: An imperative style, high-performance deep learning library}, in: Advances in Neural Information Processing Systems, Vol.~32, 2019, pp. 8024--8035.
\newline\urlprefix\url{https://proceedings.neurips.cc/paper/2019/hash/bdbca288fee7f92f2bfa9f7012727740-Abstract.html}

\bibitem{drautz2019ace}
R.~Drautz, Atomic cluster expansion for accurate and transferable interatomic potentials, Physical Review B 99 (2019) 014104.
\newblock \href {https://doi.org/10.1103/PhysRevB.99.014104} {\path{doi:10.1103/PhysRevB.99.014104}}.

\bibitem{lysogorskiy2021pace}
Y.~Lysogorskiy, C.~van~der Oord, A.~Bochkarev, S.~Menon, M.~Rinaldi, T.~Hammerschmidt, M.~Mrovec, A.~Thompson, G.~Cs{\'a}nyi, C.~Ortner, et~al., Performant implementation of the atomic cluster expansion ({PACE}) and application to copper and silicon, npj Computational Materials 7 (2021) 97.
\newblock \href {https://doi.org/10.1038/s41524-021-00559-9} {\path{doi:10.1038/s41524-021-00559-9}}.

\bibitem{bochkarev2022pacemaker}
A.~Bochkarev, Y.~Lysogorskiy, S.~Menon, M.~Qamar, M.~Mrovec, R.~Drautz, Efficient parametrization of the atomic cluster expansion, Physical Review Materials 6 (2022) 013804.
\newblock \href {https://doi.org/10.1103/PhysRevMaterials.6.013804} {\path{doi:10.1103/PhysRevMaterials.6.013804}}.

\bibitem{torchsim2025}
O.~Cohen, J.~Riebesell, R.~Goodall, A.~Kolluru, S.~Falletta, J.~Krause, J.~Colindres, G.~Ceder, A.~S. Gangan, {TorchSim}: An efficient atomistic simulation engine in {PyTorch}, AI for Science 1 (2025) 025003.
\newblock \href {https://doi.org/10.1088/3050-287X/ae1799} {\path{doi:10.1088/3050-287X/ae1799}}.

\bibitem{waasmaier1995}
D.~Waasmaier, A.~Kirfel, New analytical scattering-factor functions for free atoms and ions, Acta Crystallographica Section A 51 (1995) 416--431.
\newblock \href {https://doi.org/10.1107/S0108767394013292} {\path{doi:10.1107/S0108767394013292}}.

\bibitem{sears1992}
V.~F. Sears, Neutron scattering lengths and cross sections, Neutron News 3 (1992) 26--37.
\newblock \href {https://doi.org/10.1080/10448639208218770} {\path{doi:10.1080/10448639208218770}}.

\bibitem{yoo2003slurm}
A.~B. Yoo, M.~A. Jette, M.~Grondona, {SLURM}: Simple {Linux} utility for resource management, in: Job Scheduling Strategies for Parallel Processing, Vol. 2862 of Lecture Notes in Computer Science, Springer, 2003, pp. 44--60.
\newblock \href {https://doi.org/10.1007/10968987_3} {\path{doi:10.1007/10968987_3}}.

\bibitem{shannon1976}
R.~D. Shannon, Revised effective ionic radii and systematic studies of interatomic distances in halides and chalcogenides, Acta Crystallographica Section A 32 (1976) 751--767.
\newblock \href {https://doi.org/10.1107/S0567739476001551} {\path{doi:10.1107/S0567739476001551}}.

\bibitem{cordero2008}
B.~Cordero, V.~G{\'o}mez, A.~E. Platero-Prats, M.~Rev{\'e}s, J.~Echeverr{\'\i}a, E.~Cremades, F.~Barrag{\'a}n, S.~Alvarez, Covalent radii revisited, Dalton Transactions (2008) 2832--2838\href {https://doi.org/10.1039/b801115j} {\path{doi:10.1039/b801115j}}.

\bibitem{greenwood1997}
N.~N. Greenwood, A.~Earnshaw, Chemistry of the Elements, 2nd Edition, Butterworth-Heinemann, 1997.

\bibitem{liu2026as}
Y.~Liu, Y.~Zhou, R.~Ademuwagun, L.~Walterbos, J.~George, S.~R. Elliott, V.~L. Deringer, Medium-range structural order in amorphous arsenic, Journal of the American Chemical Society 148 (2026) 9400--9412.
\newblock \href {https://doi.org/10.1021/jacs.5c18688} {\path{doi:10.1021/jacs.5c18688}}.

\bibitem{smith1975}
P.~M. Smith, A.~J. Leadbetter, A.~J. Apling, The structures of orthorhombic and vitreous arsenic, Philosophical Magazine 31 (1975) 57--64.
\newblock \href {https://doi.org/10.1080/14786437508229285} {\path{doi:10.1080/14786437508229285}}.

\bibitem{schmidt2023adma}
J.~Schmidt, N.~Hoffmann, H.-C. Wang, P.~Borlido, P.~J. M.~A. Carri{\c{c}}o, T.~F.~T. Cerqueira, S.~Botti, M.~A.~L. Marques, Machine-learning-assisted determination of the global zero-temperature phase diagram of materials, Advanced Materials 35~(22) (2023) 2210788.
\newblock \href {https://doi.org/10.1002/adma.202210788} {\path{doi:10.1002/adma.202210788}}.

\bibitem{schmidt2024alexandria}
J.~Schmidt, T.~F.~T. Cerqueira, A.~H. Romero, A.~Loew, F.~J{\"a}ger, H.-C. Wang, S.~Botti, M.~A.~L. Marques, Improving machine-learning models in materials science through large datasets, Materials Today Physics 48 (2024) 101560.
\newblock \href {https://doi.org/10.1016/j.mtphys.2024.101560} {\path{doi:10.1016/j.mtphys.2024.101560}}.

\bibitem{barrosoluque2024omat24}
L.~Barroso-Luque, M.~Shuaibi, X.~Fu, B.~M. Wood, M.~Dzamba, M.~Gao, A.~Rizvi, M.~Uyttendaele, C.~L. Zitnick, Z.~W. Ulissi, The {Open Materials} 2024 ({OMat24}) inorganic materials dataset and models, Nature Computational Science 6 (2026) 642--652.
\newblock \href {https://doi.org/10.1038/s43588-026-00996-w} {\path{doi:10.1038/s43588-026-00996-w}}.

\bibitem{kaewmeechai2025}
C.~Kaewmeechai, J.~Strand, A.~L. Shluger, Electronic structure and properties of trapped holes in crystalline and amorphous {Ga$_2$O$_3$}, Physical Review B 111 (2025) 035203.
\newblock \href {https://doi.org/10.1103/PhysRevB.111.035203} {\path{doi:10.1103/PhysRevB.111.035203}}.

\bibitem{han2020}
Z.~Han, H.~Liang, W.~Huo, X.~Zhu, X.~Du, Z.~Mei, Boosted {UV} photodetection performance in chemically etched amorphous {Ga$_2$O$_3$} thin-film transistors, Advanced Optical Materials 8 (2020) 1901833.
\newblock \href {https://doi.org/10.1002/adom.201901833} {\path{doi:10.1002/adom.201901833}}.

\bibitem{liu2023}
Y.~Liu, H.~Liang, L.~Yang, G.~Yang, H.~Yang, S.~Song, Z.~Mei, G.~Cs{\'a}nyi, B.~Cao, Unraveling thermal transport correlated with atomistic structures in amorphous gallium oxide via machine learning combined with experiments, Advanced Materials 35 (2023) 2210873.
\newblock \href {https://doi.org/10.1002/adma.202210873} {\path{doi:10.1002/adma.202210873}}.

\bibitem{petkov1998}
V.~Petkov, G.~Holzh{\"u}ter, U.~Tr{\"o}ge, T.~Gerber, B.~Himmel, Atomic-scale structure of amorphous {TiO$_2$} by electron, {X}-ray diffraction and reverse {Monte Carlo} simulations, Journal of Non-Crystalline Solids 231 (1998) 17--30.
\newblock \href {https://doi.org/10.1016/S0022-3093(98)00418-9} {\path{doi:10.1016/S0022-3093(98)00418-9}}.

\bibitem{kumara2021}
L.~S.~R. Kumara, K.~Ishikawa, K.~Ide, H.~Hosono, T.~Kamiya, O.~Sakata, Local structure properties of hydrogenated and nonhydrogenated amorphous {In-Ga-Zn-O} thin films using {XAFS} and high-energy {XRD}, The Journal of Physical Chemistry C 125 (2021) 13619--13628.
\newblock \href {https://doi.org/10.1021/acs.jpcc.1c02437} {\path{doi:10.1021/acs.jpcc.1c02437}}.

\bibitem{mergel2000}
D.~Mergel, D.~Buschendorf, S.~Eggert, R.~Grammes, B.~Samset, Density and refractive index of {TiO$_2$} films prepared by reactive evaporation, Thin Solid Films 371 (2000) 218--224.
\newblock \href {https://doi.org/10.1016/S0040-6090(00)01015-4} {\path{doi:10.1016/S0040-6090(00)01015-4}}.

\bibitem{ide2017}
K.~Ide, M.~Kikuchi, M.~Ota, M.~Sasase, H.~Hiramatsu, H.~Kumomi, H.~Hosono, T.~Kamiya, Effects of working pressure and annealing on bulk density and nanopore structures in amorphous {In-Ga-Zn-O} thin-film transistors, Japanese Journal of Applied Physics 56 (2017) 03BB03.
\newblock \href {https://doi.org/10.7567/JJAP.56.03BB03} {\path{doi:10.7567/JJAP.56.03BB03}}.

\bibitem{hu2021}
S.~Hu, K.~Lu, H.~Ning, R.~Yao, Y.~Gong, Z.~Pan, C.~Guo, J.~Wang, C.~Pang, Z.~Gong, et~al., Study of the correlation between the amorphous indium-gallium-zinc oxide film quality and the thin-film transistor performance, Nanomaterials 11 (2021) 522.
\newblock \href {https://doi.org/10.3390/nano11020522} {\path{doi:10.3390/nano11020522}}.

\bibitem{passlack1995}
M.~Passlack, E.~F. Schubert, W.~S. Hobson, M.~Hong, N.~Moriya, S.~N.~G. Chu, K.~Konstadinidis, J.~P. Mannaerts, M.~L. Schnoes, G.~J. Zydzik, {Ga$_2$O$_3$} films for electronic and optoelectronic applications, Journal of Applied Physics 77 (1995) 686--693.
\newblock \href {https://doi.org/10.1063/1.359055} {\path{doi:10.1063/1.359055}}.

\bibitem{nieminen1996}
M.~Nieminen, L.~Niinist{\"o}, E.~Rauhala, Growth of gallium oxide thin films from gallium acetylacetonate by atomic layer epitaxy, Journal of Materials Chemistry 6 (1996) 27--31.
\newblock \href {https://doi.org/10.1039/JM9960600027} {\path{doi:10.1039/JM9960600027}}.

\bibitem{kohara2007}
S.~Kohara, M.~Itou, K.~Suzuya, Y.~Inamura, Y.~Sakurai, Y.~Ohishi, M.~Takata, Structural studies of disordered materials using high-energy {X}-ray diffraction from ambient to extreme conditions, Journal of Physics: Condensed Matter 19 (2007) 506101.
\newblock \href {https://doi.org/10.1088/0953-8984/19/50/506101} {\path{doi:10.1088/0953-8984/19/50/506101}}.

\bibitem{yang2018tio2}
K.~Yang, A.~Kachmar, B.~Wang, N.~M.~A. Krishnan, M.~Balonis, G.~Sant, M.~Bauchy, New insights into the atomic structure of amorphous {TiO$_2$} using tight-binding molecular dynamics, The Journal of Chemical Physics 149 (2018) 094501.
\newblock \href {https://doi.org/10.1063/1.5042783} {\path{doi:10.1063/1.5042783}}.

\bibitem{zeidler2014}
A.~Zeidler, K.~Wezka, R.~F. Rowlands, D.~A.~J. Whittaker, P.~S. Salmon, A.~Polidori, J.~W.~E. Drewitt, S.~Klotz, H.~E. Fischer, M.~C. Wilding, et~al., High-pressure transformation of {SiO$_2$} glass from a tetrahedral to an octahedral network: a joint approach using neutron diffraction and molecular dynamics, Physical Review Letters 113 (2014) 135501.
\newblock \href {https://doi.org/10.1103/PhysRevLett.113.135501} {\path{doi:10.1103/PhysRevLett.113.135501}}.

\bibitem{goetzke1991}
K.~Goetzke, H.-J. Klein, Properties and efficient algorithmic determination of different classes of rings in finite and infinite polyhedral networks, Journal of Non-Crystalline Solids 127 (1991) 215--220.
\newblock \href {https://doi.org/10.1016/0022-3093(91)90145-V} {\path{doi:10.1016/0022-3093(91)90145-V}}.

\bibitem{yuan2002}
X.~Yuan, A.~N. Cormack, Efficient algorithm for primitive ring statistics in topological networks, Computational Materials Science 24 (2002) 343--360.
\newblock \href {https://doi.org/10.1016/S0927-0256(01)00256-7} {\path{doi:10.1016/S0927-0256(01)00256-7}}.

\bibitem{mozzi1969}
R.~L. Mozzi, B.~E. Warren, The structure of vitreous silica, Journal of Applied Crystallography 2 (1969) 164--172.
\newblock \href {https://doi.org/10.1107/S0021889869006868} {\path{doi:10.1107/S0021889869006868}}.

\bibitem{wright1994}
A.~C. Wright, Neutron scattering from vitreous silica. {V}. {The} structure of vitreous silica: What have we learned from 60 years of diffraction studies?, Journal of Non-Crystalline Solids 179 (1994) 84--115.
\newblock \href {https://doi.org/10.1016/0022-3093(94)90687-4} {\path{doi:10.1016/0022-3093(94)90687-4}}.

\bibitem{vollmayr1996}
K.~Vollmayr, W.~Kob, K.~Binder, Cooling-rate effects in amorphous silica: a computer-simulation study, Physical Review B 54 (1996) 15808--15827.
\newblock \href {https://doi.org/10.1103/PhysRevB.54.15808} {\path{doi:10.1103/PhysRevB.54.15808}}.

\bibitem{feff85exafs}
{FEFF Project and xraypy developers}, \href{https://github.com/xraypy/feff85exafs}{{Feff8L} (feff85exafs): open-source theoretical {EXAFS} calculations}, based on {FEFF8.5L}; revision ec8dcb07ca8ee034d0fa7431782074f0f65357a5 (2026).
\newline\urlprefix\url{https://github.com/xraypy/feff85exafs}

\bibitem{revenant2015}
C.~Revenant, M.~Benwadih, O.~Proux, Local structure around {Zn} and {Ga} in solution-processed {In-Ga-Zn-O} and implications for electronic properties, physica status solidi (RRL) -- Rapid Research Letters 9 (2015) 652--655.
\newblock \href {https://doi.org/10.1002/pssr.201510322} {\path{doi:10.1002/pssr.201510322}}.

\bibitem{morganHofmann}
B.~J. Morgan, \href{https://github.com/bjmorgan/hofmann}{hofmann}, version 0.21.1, revision 87627da; accessed 13 November 2026 (2026).
\newline\urlprefix\url{https://github.com/bjmorgan/hofmann}

\end{thebibliography}


\begin{thebibliography}{10}
\expandafter\ifx\csname url\endcsname\relax
  \def\url#1{\texttt{#1}}\fi
\expandafter\ifx\csname urlprefix\endcsname\relax\def\urlprefix{URL }\fi
\expandafter\ifx\csname href\endcsname\relax
  \def\href#1#2{#2} \def\path#1{#1}\fi

\bibitem{pauling1932}
L.~Pauling, The nature of the chemical bond. {IV}. {The} energy of single bonds and the relative electronegativity of atoms, Journal of the American Chemical Society 54 (1932) 3570--3582.
\newblock \href {https://doi.org/10.1021/ja01348a011} {\path{doi:10.1021/ja01348a011}}.

\bibitem{pauling1960}
L.~Pauling, The Nature of the Chemical Bond and the Structure of Molecules and Crystals: An Introduction to Modern Structural Chemistry, 3rd Edition, Cornell University Press, Ithaca, NY, 1960.

\bibitem{allred1961}
A.~L. Allred, Electronegativity values from thermochemical data, Journal of Inorganic and Nuclear Chemistry 17 (1961) 215--221.
\newblock \href {https://doi.org/10.1016/0022-1902(61)80142-5} {\path{doi:10.1016/0022-1902(61)80142-5}}.

\bibitem{shannon1976}
R.~D. Shannon, Revised effective ionic radii and systematic studies of interatomic distances in halides and chalcogenides, Acta Crystallographica Section A 32 (1976) 751--767.
\newblock \href {https://doi.org/10.1107/S0567739476001551} {\path{doi:10.1107/S0567739476001551}}.

\bibitem{cordero2008}
B.~Cordero, V.~G{\'o}mez, A.~E. Platero-Prats, M.~Rev{\'e}s, J.~Echeverr{\'\i}a, E.~Cremades, F.~Barrag{\'a}n, S.~Alvarez, Covalent radii revisited, Dalton Transactions (2008) 2832--2838\href {https://doi.org/10.1039/b801115j} {\path{doi:10.1039/b801115j}}.

\bibitem{greenwood1997}
N.~N. Greenwood, A.~Earnshaw, Chemistry of the Elements, 2nd Edition, Butterworth-Heinemann, 1997.

\bibitem{youn2014}
Y.~Youn, Y.~Kang, S.~Han, An efficient method to generate amorphous structures based on local geometry, Computational Materials Science 95 (2014) 256--262.
\newblock \href {https://doi.org/10.1016/j.commatsci.2014.07.053} {\path{doi:10.1016/j.commatsci.2014.07.053}}.

\bibitem{haynes2016crc}
W.~M. Haynes, D.~R. Lide, T.~J. Bruno (Eds.), {CRC} Handbook of Chemistry and Physics, 97th Edition, CRC Press, Boca Raton, FL, 2016.
\newblock \href {https://doi.org/10.1201/9781315380476} {\path{doi:10.1201/9781315380476}}.

\bibitem{pubchem_as2o3}
{National Center for Biotechnology Information}, \href{https://pubchem.ncbi.nlm.nih.gov/compound/14888}{{PubChem} compound summary for {CID} 14888, arsenic trioxide}, accessed October 2025 (2025).
\newline\urlprefix\url{https://pubchem.ncbi.nlm.nih.gov/compound/14888}

\bibitem{napolitano1965}
A.~Napolitano, P.~B. Macedo, E.~G. Hawkins, Viscosity and density of boron trioxide, Journal of the American Ceramic Society 48 (1965) 613--616.
\newblock \href {https://doi.org/10.1111/j.1151-2916.1965.tb14690.x} {\path{doi:10.1111/j.1151-2916.1965.tb14690.x}}.

\bibitem{micoulaut2006}
M.~Micoulaut, L.~Cormier, G.~S. Henderson, The structure of amorphous, crystalline and liquid {GeO$_2$}, Journal of Physics: Condensed Matter 18 (2006) R753--R784.
\newblock \href {https://doi.org/10.1088/0953-8984/18/45/R01} {\path{doi:10.1088/0953-8984/18/45/R01}}.

\bibitem{bruckner1970}
R.~Br{\"u}ckner, Properties and structure of vitreous silica. {I}, Journal of Non-Crystalline Solids 5 (1970) 123--175.
\newblock \href {https://doi.org/10.1016/0022-3093(70)90190-0} {\path{doi:10.1016/0022-3093(70)90190-0}}.

\bibitem{alderman2020teo2}
O.~L.~G. Alderman, C.~J. Benmore, S.~Feller, E.~I. Kamitsos, E.~D. Simandiras, D.~G. Liakos, M.~Jesuit, M.~Boyd, M.~Packard, R.~Weber, Short-range disorder in {TeO$_2$} melt and glass, The Journal of Physical Chemistry Letters 11 (2020) 427--431.
\newblock \href {https://doi.org/10.1021/acs.jpclett.9b03231} {\path{doi:10.1021/acs.jpclett.9b03231}}.

\bibitem{pugliese2022}
A.~Pugliese, B.~Shyam, G.~M. Repa, A.~H. Nguyen, A.~Mehta, E.~B. Webb, III, L.~A. Fredin, N.~C. Strandwitz, Atomic-layer-deposited aluminum oxide thin films probed with {X}-ray scattering and compared to molecular dynamics and density functional theory models, ACS Omega 7 (2022) 41033--41043.
\newblock \href {https://doi.org/10.1021/acsomega.2c04402} {\path{doi:10.1021/acsomega.2c04402}}.

\bibitem{liu2023}
Y.~Liu, H.~Liang, L.~Yang, G.~Yang, H.~Yang, S.~Song, Z.~Mei, G.~Cs{\'a}nyi, B.~Cao, Unraveling thermal transport correlated with atomistic structures in amorphous gallium oxide via machine learning combined with experiments, Advanced Materials 35 (2023) 2210873.
\newblock \href {https://doi.org/10.1002/adma.202210873} {\path{doi:10.1002/adma.202210873}}.

\bibitem{kumara2021}
L.~S.~R. Kumara, K.~Ishikawa, K.~Ide, H.~Hosono, T.~Kamiya, O.~Sakata, Local structure properties of hydrogenated and nonhydrogenated amorphous {In-Ga-Zn-O} thin films using {XAFS} and high-energy {XRD}, The Journal of Physical Chemistry C 125 (2021) 13619--13628.
\newblock \href {https://doi.org/10.1021/acs.jpcc.1c02437} {\path{doi:10.1021/acs.jpcc.1c02437}}.

\bibitem{zavyalova1976}
A.~A. Zav'yalova, R.~M. Imamov, N.~A. Ragimli, S.~A. Semiletov, Electron-diffraction study of the structure of cubic {C-Sm$_2$O$_3$}, Kristallografiya 21 (1976) 727--729.

\bibitem{mergel2000}
D.~Mergel, D.~Buschendorf, S.~Eggert, R.~Grammes, B.~Samset, Density and refractive index of {TiO$_2$} films prepared by reactive evaporation, Thin Solid Films 371 (2000) 218--224.
\newblock \href {https://doi.org/10.1016/S0040-6090(00)01015-4} {\path{doi:10.1016/S0040-6090(00)01015-4}}.

\bibitem{kim2020hfo2}
K.-M. Kim, J.~S. Jang, S.-G. Yoon, J.-Y. Yun, N.-K. Chung, Structural, optical and electrical properties of {HfO$_2$} thin films deposited at low-temperature using plasma-enhanced atomic layer deposition, Materials 13 (2020) 2008.
\newblock \href {https://doi.org/10.3390/ma13092008} {\path{doi:10.3390/ma13092008}}.

\bibitem{shank2018}
J.~C. Shank, M.~B. Tellekamp, M.~J. Wahila, S.~Howard, A.~S. Weidenbach, B.~Zivasatienraj, L.~F.~J. Piper, W.~A. Doolittle, Scalable memdiodes exhibiting rectification and hysteresis for neuromorphic computing, Scientific Reports 8 (2018) 12935.
\newblock \href {https://doi.org/10.1038/s41598-018-30727-9} {\path{doi:10.1038/s41598-018-30727-9}}.

\bibitem{alderman2018tantala}
O.~L.~G. Alderman, C.~J. Benmore, J.~Neuefeind, E.~Coillet, A.~Mermet, V.~Martinez, A.~Tamalonis, R.~Weber, Amorphous tantala and its relationship with the molten state, Physical Review Materials 2 (2018) 043602.
\newblock \href {https://doi.org/10.1103/PhysRevMaterials.2.043602} {\path{doi:10.1103/PhysRevMaterials.2.043602}}.

\bibitem{ketterer1986}
J.~Ketterer, V.~Kr{\"a}mer, Structure refinement of bismuth oxide bromide, {BiOBr}, Acta Crystallographica Section C 42 (1986) 1098--1099.
\newblock \href {https://doi.org/10.1107/S0108270186093307} {\path{doi:10.1107/S0108270186093307}}.

\bibitem{handbook_mineralogy_bismoclite}
J.~W. Anthony, R.~A. Bideaux, K.~W. Bladh, M.~C. Nichols, \href{https://www.handbookofmineralogy.org/pdfs/bismoclite.pdf}{{Handbook of Mineralogy}: Bismoclite, {BiOCl}}, accessed 13 November 2026 (2005).
\newline\urlprefix\url{https://www.handbookofmineralogy.org/pdfs/bismoclite.pdf}

\bibitem{lind1970}
M.~D. Lind, Refinement of the crystal structure of iron oxychloride, Acta Crystallographica Section B 26 (1970) 1058--1062.
\newblock \href {https://doi.org/10.1107/S0567740870003618} {\path{doi:10.1107/S0567740870003618}}.

\bibitem{maslen1996}
E.~N. Maslen, V.~A. Streltsov, N.~R. Streltsova, N.~Ishizawa, Synchrotron {X}-ray electron density in the layered {LaOCl} structure, Acta Crystallographica Section B 52 (1996) 576--579.
\newblock \href {https://doi.org/10.1107/S0108768196003084} {\path{doi:10.1107/S0108768196003084}}.

\bibitem{wright1989}
A.~C. Wright, A.~G. Clare, G.~Etherington, R.~N. Sinclair, S.~A. Brawer, M.~J. Weber, A neutron diffraction and molecular dynamics investigation of the structure of vitreous beryllium fluoride, Journal of Non-Crystalline Solids 111 (1989) 139--152.
\newblock \href {https://doi.org/10.1016/0022-3093(89)90275-5} {\path{doi:10.1016/0022-3093(89)90275-5}}.

\bibitem{desa1982}
J.~A. Erwin~Desa, A.~C. Wright, J.~Wong, R.~N. Sinclair, A neutron diffraction investigation of the structure of vitreous zinc chloride, Journal of Non-Crystalline Solids 51 (1982) 57--86.
\newblock \href {https://doi.org/10.1016/0022-3093(82)90189-2} {\path{doi:10.1016/0022-3093(82)90189-2}}.

\bibitem{seo2004}
H.-S. Seo, T.-Y. Lee, J.~G. Wen, I.~Petrov, J.~E. Greene, D.~Gall, Growth and physical properties of epitaxial {HfN} layers on {MgO(001)}, Journal of Applied Physics 96 (2004) 878--884.
\newblock \href {https://doi.org/10.1063/1.1759783} {\path{doi:10.1063/1.1759783}}.

\bibitem{gall2001}
D.~Gall, M.~St{\"a}dele, K.~J{\"a}rrendahl, I.~Petrov, P.~Desjardins, R.~T. Haasch, T.-Y. Lee, J.~E. Greene, Electronic structure of {ScN} determined using optical spectroscopy, photoemission, and ab initio calculations, Physical Review B 63 (2001) 125119.
\newblock \href {https://doi.org/10.1103/PhysRevB.63.125119} {\path{doi:10.1103/PhysRevB.63.125119}}.

\bibitem{du2010}
L.~Du, J.~H. Edgar, R.~A. Peascoe-Meisner, Y.~Gong, S.~Bakalova, M.~Kuball, Sublimation crystal growth of yttrium nitride, Journal of Crystal Growth 312 (2010) 2896--2903.
\newblock \href {https://doi.org/10.1016/j.jcrysgro.2010.06.011} {\path{doi:10.1016/j.jcrysgro.2010.06.011}}.

\bibitem{sattari2025}
S.~M. Sattari-Esfahlan, S.~Mirzaei, M.~J. Josline, J.-Y. Moon, S.-H. Hyun, H.~Jang, J.-H. Lee, Amorphous boron nitride: synthesis, properties and device application, Nano Convergence 12 (2025) 22.
\newblock \href {https://doi.org/10.1186/s40580-025-00486-1} {\path{doi:10.1186/s40580-025-00486-1}}.

\bibitem{ruddlesden1958}
S.~N. Ruddlesden, P.~Popper, On the crystal structure of the nitrides of silicon and germanium, Acta Crystallographica 11 (1958) 465--468.
\newblock \href {https://doi.org/10.1107/S0365110X58001304} {\path{doi:10.1107/S0365110X58001304}}.

\bibitem{yen2003}
B.~K. Yen, R.~L. White, R.~J. Waltman, Q.~Dai, D.~C. Miller, A.~J. Kellock, B.~Marchon, P.~H. Kasai, M.~F. Toney, B.~R. York, et~al., Microstructure and properties of ultrathin amorphous silicon nitride protective coating, Journal of Vacuum Science \& Technology A 21 (2003) 1895--1904.
\newblock \href {https://doi.org/10.1116/1.1615974} {\path{doi:10.1116/1.1615974}}.

\bibitem{cornet1973}
J.~Cornet, D.~Rossier, Properties and structure of {As--Te} glasses: ({I}). {Glass-forming} ability and related properties, Journal of Non-Crystalline Solids 12 (1973) 61--84.
\newblock \href {https://doi.org/10.1016/0022-3093(73)90055-0} {\path{doi:10.1016/0022-3093(73)90055-0}}.

\bibitem{shah2021}
N.~A. Shah, W.~Mahmood, M.~Abbas, N.~Nazar, A.~H. Khosa, A.~Zeb, A.~Malik, The synthesis of {CdZnTe} semiconductor thin films for tandem solar cells, RSC Advances 11 (2021) 39940--39949.
\newblock \href {https://doi.org/10.1039/D1RA07755D} {\path{doi:10.1039/D1RA07755D}}.

\bibitem{njoroge2002}
W.~K. Njoroge, H.-W. W{\"o}ltgens, M.~Wuttig, Density changes upon crystallization of {Ge$_2$Sb$_{2.04}$Te$_{4.74}$} films, Journal of Vacuum Science \& Technology A 20 (2002) 230--233.
\newblock \href {https://doi.org/10.1116/1.1430249} {\path{doi:10.1116/1.1430249}}.

\bibitem{ghezzi2011}
G.~E. Ghezzi, J.~Y. Raty, S.~Maitrejean, A.~Roule, E.~Elkaim, F.~Hippert, Effect of carbon doping on the structure of amorphous {GeTe} phase change material, Applied Physics Letters 99 (2011) 151906.
\newblock \href {https://doi.org/10.1063/1.3651321} {\path{doi:10.1063/1.3651321}}.

\bibitem{madelung2004}
O.~Madelung, Semiconductors: Data Handbook, 3rd Edition, Springer, Berlin, 2004.
\newblock \href {https://doi.org/10.1007/978-3-642-18865-7} {\path{doi:10.1007/978-3-642-18865-7}}.

\bibitem{knotek2020}
P.~Knotek, P.~Kut{\'a}lek, E.~{\v{C}}erno{\v{s}}kov{\'a}, M.~Vl{\v{c}}ek, L.~Tich{\'y}, The density, nanohardness and some optical properties of {As--S} and {As--Se} chalcogenide bulk glasses and thin films, RSC Advances 10 (2020) 42744--42753.
\newblock \href {https://doi.org/10.1039/D0RA08939G} {\path{doi:10.1039/D0RA08939G}}.

\bibitem{tverjanovich2021}
A.~Tverjanovich, M.~Khomenko, C.~J. Benmore, M.~Bokova, A.~Sokolov, D.~Fontanari, M.~Kassem, T.~Usuki, E.~Bychkov, Bulk glassy {GeTe$_2$}: a missing member of the tetrahedral {GeX$_2$} family and a precursor for the next generation of phase-change materials, Chemistry of Materials 33 (2021) 1031--1045.
\newblock \href {https://doi.org/10.1021/acs.chemmater.0c04409} {\path{doi:10.1021/acs.chemmater.0c04409}}.

\bibitem{putero2014}
M.~Putero, M.-V. Coulet, C.~Muller, G.~Cohen, M.~Hopstaken, C.~Baehtz, S.~Raoux, Density change upon crystallization of {Ga--Sb} films, Applied Physics Letters 105 (2014) 181910.
\newblock \href {https://doi.org/10.1063/1.4901321} {\path{doi:10.1063/1.4901321}}.

\bibitem{cliche1994}
L.~Cliche, S.~Roorda, R.~A. Masut, Persistent room-temperature relaxation of {InP} amorphized and compacted by {MeV} ion beams, Applied Physics Letters 65 (1994) 1754--1756.
\newblock \href {https://doi.org/10.1063/1.112907} {\path{doi:10.1063/1.112907}}.

\bibitem{donovan1970}
T.~M. Donovan, W.~E. Spicer, J.~M. Bennett, E.~J. Ashley, Optical properties of amorphous germanium films, Physical Review B 2 (1970) 397--413.
\newblock \href {https://doi.org/10.1103/PhysRevB.2.397} {\path{doi:10.1103/PhysRevB.2.397}}.

\bibitem{custer1994}
J.~S. Custer, M.~O. Thompson, D.~C. Jacobson, J.~M. Poate, S.~Roorda, W.~C. Sinke, F.~Spaepen, Density of amorphous {Si}, Applied Physics Letters 64 (1994) 437--439.
\newblock \href {https://doi.org/10.1063/1.111121} {\path{doi:10.1063/1.111121}}.

\bibitem{smith1975}
P.~M. Smith, A.~J. Leadbetter, A.~J. Apling, The structures of orthorhombic and vitreous arsenic, Philosophical Magazine 31 (1975) 57--64.
\newblock \href {https://doi.org/10.1080/14786437508229285} {\path{doi:10.1080/14786437508229285}}.

\bibitem{libassi2000}
A.~LiBassi, A.~C. Ferrari, V.~Stolojan, B.~K. Tanner, J.~Robertson, L.~M. Brown, Density, {sp$^3$} content and internal layering of {DLC} films by {X}-ray reflectivity and electron energy loss spectroscopy, Diamond and Related Materials 9 (2000) 771--776.
\newblock \href {https://doi.org/10.1016/S0925-9635(99)00233-2} {\path{doi:10.1016/S0925-9635(99)00233-2}}.

\bibitem{remes1998}
Z.~Reme{\v{s}}, M.~Van{\v{e}}{\v{c}}ek, P.~Torres, U.~Kroll, A.~H. Mahan, R.~S. Crandall, Optical determination of the mass density of amorphous and microcrystalline silicon layers with different hydrogen contents, Journal of Non-Crystalline Solids 227--230 (1998) 876--879.
\newblock \href {https://doi.org/10.1016/S0022-3093(98)00207-5} {\path{doi:10.1016/S0022-3093(98)00207-5}}.

\bibitem{campbell2023}
A.~Campbell, E.~Cakmak, B.~Henry, K.~Johnson, D.~Muzquiz, Y.~Osetskiy, E.~Paxton, S.~Raiman, D.~Sulejmanovic, E.~Zarkadoula, et~al., {Be$_2$C} synthesis, properties, and ion-beam irradiation damage characterization, Tech. Rep. ORNL/TM-2023/3011, Oak Ridge National Laboratory (2023).
\newblock \href {https://doi.org/10.2172/1997705} {\path{doi:10.2172/1997705}}.

\bibitem{heera1997}
V.~Heera, F.~Prokert, N.~Schell, H.~Seifarth, W.~Fukarek, M.~Voelskow, W.~Skorupa, Density and structural changes in {SiC} after amorphization and annealing, Applied Physics Letters 70 (1997) 3531--3533.
\newblock \href {https://doi.org/10.1063/1.119223} {\path{doi:10.1063/1.119223}}.

\bibitem{wong2018}
K.~Wong, R.~P. Krishnan, E.~M. Dufresne, K.~Ohara, A.~R. Sandy, S.~M. Chathoth, Universal aging characteristics of macroscopically and microscopically dissimilar metallic glasses, Acta Materialia 155 (2018) 35--42.
\newblock \href {https://doi.org/10.1016/j.actamat.2018.05.059} {\path{doi:10.1016/j.actamat.2018.05.059}}.

\bibitem{hasegawa1978}
R.~Hasegawa, R.~Ray, Iron-boron metallic glasses, Journal of Applied Physics 49 (1978) 4174--4179.
\newblock \href {https://doi.org/10.1063/1.325328} {\path{doi:10.1063/1.325328}}.

\bibitem{basak1979}
S.~Basak, R.~Clarke, S.~R. Nagel, Pair distribution function and its relation to the glass transition in an amorphous alloy, Physical Review B 20 (1979) 3388--3390.
\newblock \href {https://doi.org/10.1103/PhysRevB.20.3388} {\path{doi:10.1103/PhysRevB.20.3388}}.

\bibitem{otsuka2005}
K.~Otsuka, X.~Ren, Physical metallurgy of {Ti--Ni}-based shape memory alloys, Progress in Materials Science 50 (2005) 511--678.
\newblock \href {https://doi.org/10.1016/j.pmatsci.2004.10.001} {\path{doi:10.1016/j.pmatsci.2004.10.001}}.

\bibitem{polk1972}
D.~E. Polk, The structure of glassy metallic alloys, Acta Metallurgica 20 (1972) 485--491.
\newblock \href {https://doi.org/10.1016/0001-6160(72)90003-X} {\path{doi:10.1016/0001-6160(72)90003-X}}.

\bibitem{larsen2017}
A.~{Hjorth Larsen}, J.~J. Mortensen, J.~Blomqvist, I.~E. Castelli, R.~Christensen, M.~Du{\l}ak, J.~Friis, M.~N. Groves, B.~Hammer, C.~Hargus, et~al., The atomic simulation environment --- a {Python} library for working with atoms, Journal of Physics: Condensed Matter 29 (2017) 273002.
\newblock \href {https://doi.org/10.1088/1361-648X/aa680e} {\path{doi:10.1088/1361-648X/aa680e}}.

\bibitem{torchsim2025}
O.~Cohen, J.~Riebesell, R.~Goodall, A.~Kolluru, S.~Falletta, J.~Krause, J.~Colindres, G.~Ceder, A.~S. Gangan, {TorchSim}: An efficient atomistic simulation engine in {PyTorch}, AI for Science 1 (2025) 025003.
\newblock \href {https://doi.org/10.1088/3050-287X/ae1799} {\path{doi:10.1088/3050-287X/ae1799}}.

\bibitem{ewald1921}
P.~P. Ewald, Die {Berechnung} optischer und elektrostatischer {Gitterpotentiale}, Annalen der Physik 369 (1921) 253--287.
\newblock \href {https://doi.org/10.1002/andp.19213690304} {\path{doi:10.1002/andp.19213690304}}.

\bibitem{liu2026as}
Y.~Liu, Y.~Zhou, R.~Ademuwagun, L.~Walterbos, J.~George, S.~R. Elliott, V.~L. Deringer, Medium-range structural order in amorphous arsenic, Journal of the American Chemical Society 148 (2026) 9400--9412.
\newblock \href {https://doi.org/10.1021/jacs.5c18688} {\path{doi:10.1021/jacs.5c18688}}.

\bibitem{faber1965}
T.~E. Faber, J.~M. Ziman, A theory of the electrical properties of liquid metals. {III}. {The} resistivity of binary alloys, Philosophical Magazine 11 (1965) 153--173.
\newblock \href {https://doi.org/10.1080/14786436508211931} {\path{doi:10.1080/14786436508211931}}.

\bibitem{waasmaier1995}
D.~Waasmaier, A.~Kirfel, New analytical scattering-factor functions for free atoms and ions, Acta Crystallographica Section A 51 (1995) 416--431.
\newblock \href {https://doi.org/10.1107/S0108767394013292} {\path{doi:10.1107/S0108767394013292}}.

\bibitem{kaewmeechai2025}
C.~Kaewmeechai, J.~Strand, A.~L. Shluger, Electronic structure and properties of trapped holes in crystalline and amorphous {Ga$_2$O$_3$}, Physical Review B 111 (2025) 035203.
\newblock \href {https://doi.org/10.1103/PhysRevB.111.035203} {\path{doi:10.1103/PhysRevB.111.035203}}.

\bibitem{blanco2005energetics}
M.~A. Blanco, M.~B. Sahariah, H.~Jiang, A.~Costales, R.~Pandey, Energetics and migration of point defects in {Ga$_2$O$_3$}, Physical Review B 72 (2005) 184103.
\newblock \href {https://doi.org/10.1103/PhysRevB.72.184103} {\path{doi:10.1103/PhysRevB.72.184103}}.

\bibitem{petkov1998}
V.~Petkov, G.~Holzh{\"u}ter, U.~Tr{\"o}ge, T.~Gerber, B.~Himmel, Atomic-scale structure of amorphous {TiO$_2$} by electron, {X}-ray diffraction and reverse {Monte Carlo} simulations, Journal of Non-Crystalline Solids 231 (1998) 17--30.
\newblock \href {https://doi.org/10.1016/S0022-3093(98)00418-9} {\path{doi:10.1016/S0022-3093(98)00418-9}}.

\bibitem{haas2009lattice}
P.~Haas, F.~Tran, P.~Blaha, Calculation of the lattice constant of solids with semilocal functionals, Physical Review B 79 (2009) 085104.
\newblock \href {https://doi.org/10.1103/PhysRevB.79.085104} {\path{doi:10.1103/PhysRevB.79.085104}}.

\bibitem{vollmayr1996}
K.~Vollmayr, W.~Kob, K.~Binder, Cooling-rate effects in amorphous silica: a computer-simulation study, Physical Review B 54 (1996) 15808--15827.
\newblock \href {https://doi.org/10.1103/PhysRevB.54.15808} {\path{doi:10.1103/PhysRevB.54.15808}}.

\bibitem{elliott1991fsdp}
S.~R. Elliott, Medium-range structural order in covalent amorphous solids, Nature 354 (1991) 445--452.
\newblock \href {https://doi.org/10.1038/354445a0} {\path{doi:10.1038/354445a0}}.

\bibitem{feff85exafs}
{FEFF Project and xraypy developers}, \href{https://github.com/xraypy/feff85exafs}{{Feff8L} (feff85exafs): open-source theoretical {EXAFS} calculations}, based on {FEFF8.5L}; revision ec8dcb07ca8ee034d0fa7431782074f0f65357a5 (2026).
\newline\urlprefix\url{https://github.com/xraypy/feff85exafs}

\bibitem{reis2023}
A.~Reis, M.~Hanke, O.~Bierwagen, A.~Trampert, P.~Mazzolini, E.~Welter, Disorder--order transition in {Ga$_2$O$_3$} and its solid solution with {In$_2$O$_3$} upon thermal annealing, physica status solidi (b) 260 (2023) 2200535.
\newblock \href {https://doi.org/10.1002/pssb.202200535} {\path{doi:10.1002/pssb.202200535}}.

\bibitem{ahman1996}
J.~{\AA}hman, G.~Svensson, J.~Albertsson, A reinvestigation of {$\beta$}-gallium oxide, Acta Crystallographica Section C 52 (1996) 1336--1338.
\newblock \href {https://doi.org/10.1107/S0108270195016404} {\path{doi:10.1107/S0108270195016404}}.

\bibitem{jia2018}
J.~Jia, A.~Suko, Y.~Shigesato, T.~Okajima, K.~Inoue, H.~Hosomi, Evolution of defect structures and deep subgap states during annealing of amorphous {In-Ga-Zn} oxide for thin-film transistors, Physical Review Applied 9 (2018) 014018.
\newblock \href {https://doi.org/10.1103/PhysRevApplied.9.014018} {\path{doi:10.1103/PhysRevApplied.9.014018}}.

\bibitem{revenant2015}
C.~Revenant, M.~Benwadih, O.~Proux, Local structure around {Zn} and {Ga} in solution-processed {In-Ga-Zn-O} and implications for electronic properties, physica status solidi (RRL) -- Rapid Research Letters 9 (2015) 652--655.
\newblock \href {https://doi.org/10.1002/pssr.201510322} {\path{doi:10.1002/pssr.201510322}}.

\end{thebibliography}

\end{document}


\begin{frontmatter}

\title{Supplementary material to: AmorphGen: A Python package for
  amorphous structure generation with machine-learning and classical
  interatomic potentials}

\author[label1]{Chaiyawat Kaewmeechai\,\orcidlink{0000-0003-1603-2659}\corref{cor1}}
\ead{c.kaewmeechai@bham.ac.uk}
\author[label1]{Louie Slocombe\,\orcidlink{0000-0002-6986-5526}}
\author[label1]{Alexis G. Manche\,\orcidlink{0000-0002-8505-474X}}
\author[label1]{David O. Scanlon\,\orcidlink{0000-0001-9174-8601}\corref{cor1}}
\ead{d.o.scanlon@bham.ac.uk}
\cortext[cor1]{Corresponding authors}
\address[label1]{School of Chemistry, University of Birmingham, Edgbaston, Birmingham B15 2TT, United Kingdom}

\end{frontmatter}

\noindent
This supplement documents the methods and checks supporting the main
paper. Sections~\ref{sec:composition}--\ref{sec:packing-calibration}
describe the composition rules and density calibration, including the
reference data and fitted packing factors.
Sections~\ref{sec:provenance}--\ref{sec:cost} record the structural
ensembles, simulation settings, analysis methods, additional structural
comparisons, and computational cost. Section~\ref{sec:exafs} presents the EXAFS protocol, validation
against FEFF reference calculations, convergence tests, and comparisons
with experiment.

\tableofcontents
\clearpage

\section{Composition rules}
\label{sec:composition}

AmorphGen classifies each element pair $A$--$B$ as ionic, covalent,
or metallic by an element-type rule, and pairs that the
rule calls ionic but whose Pauling electronegativity difference is
below 1.0 are treated as covalent
\cite{pauling1932,pauling1960,allred1961}. The minimum separation of
the pair is $d_{\min}(A,B) = (r_A + r_B)\,s_{\text{bond}}$, where $r_A$
and $r_B$ are Shannon ionic \cite{shannon1976}, Cordero covalent
\cite{cordero2008}, or Goldschmidt metallic \cite{greenwood1997} radii;
covalent pairs that include a metalloid use its metallic radius where
one is tabulated. The bond-type scale factor $s_{\text{bond}}$ is 0.80 for ionic,
covalent, and anion--anion pairs, 0.70 for large anions such as Cl and
S, and 0.85 for metal--metal pairs, and $d_{\min}$ is capped at
\SIrange{2.8}{3.0}{\angstrom}. In a compound with an anion $X$, pairs
of metal cations and pairs of the same metalloid use the edge-sharing
distance instead, $\sqrt{2}\,(r_A + r_X)\,s$, the separation of two
cations whose polyhedra share an edge. Here $r_X$ is the largest anion
radius, the cation--anion contact is averaged over the two cations of
an unlike pair, and $s = 0.85$ for metal cations and 0.80 for
metalloids. Unlike cation pairs that include a metalloid, such as
Si--Zn in \ce{Zn2SiO4}, are treated as covalent. The separation is
never smaller than the pair's metallic or covalent contact. Compounds
with many cations per anion use shorter contacts: the factor
$\sqrt{2}$ becomes 1.155 (cations \ang{70.5} apart about the anion)
for seven or eight cation neighbors per anion and 1 for nine to
twelve (\ce{Li2O}, \ce{Li3N}), and the metallic contact is used above
twelve (\ce{Fe3C}).
The class also sets coordination targets. In oxides, the tetrahedral
formers (Si, Ge, Zn, B, Be, and the other metalloids) are four-fold,
and other cations are five-fold, or six-fold in rutile-type dioxides
and high-valent oxides, and eight-fold in fluorite-type dioxides. The
oxide targets have a tolerance of one when a metal or high-valent cation
is present and none otherwise (\ce{SiO2}, \ce{GeO2}, \ce{ZnO}). Cations
in halides are six-fold, or four-fold for the tetrahedral formers
(\ce{BeF2}, \ce{ZnCl2}), and cations in nitrides are four-fold, all
with no tolerance. Most other classes also use exact targets; alloys
and pure metals are eight-fold with a tolerance of two.
The random generator then uses the seed-coordinate scheme
\cite{youn2014}. Each new atom is tried at a distance
$d_{\min} \le d \le d_{\max}$ from a seed atom that can still accept a
bond, where $d_{\max}$ is the bonding cutoff. The position is rejected
if any pair is closer than its $d_{\min}$ or a neighbor would become
over-coordinated. A repair pass moves atoms below a coordination floor
(two for anions, three for cations) next to an atom that can still
accept a bond. If placement stalls, the atoms are repacked in the same
cell, and the cell is enlarged only if repacking fails.

\begin{table}[H]
\centering
\caption{Material classes, radius tables, and packing factors of the density estimate.}
\label{tab:packing}
\footnotesize
\begin{tabular}{>{\raggedright\arraybackslash}p{0.22\textwidth}>{\raggedright\arraybackslash}p{0.15\textwidth}c>{\raggedright\arraybackslash}p{0.32\textwidth}}
\toprule
Class & Radius & $f_{\text{pack}}$ & Examples \\
\midrule
Network-forming oxide     & Shannon   & 0.50 & \ce{SiO2}, \ce{GeO2}, \ce{B2O3} \\
Metal oxide               & Shannon   & 0.52 & \ce{Ga2O3}, \ce{In2O3}, \ce{InGaZnO4}, \ce{Al2O3} \\
Rutile-type dioxide       & Shannon   & 0.66 & \ce{TiO2}, \ce{SnO2}, \ce{IrO2} \\
Fluorite-type dioxide     & Shannon   & 0.63 & \ce{HfO2}, \ce{ZrO2} \\
High-valent oxide         & Shannon   & 0.60 & \ce{Sb2O5}, \ce{Ta2O5}, \ce{WO3} \\
Small-cation oxide        & Cordero   & 0.35 & \ce{BeO} \\
Oxyhalide                 & Shannon   & interpolated\textsuperscript{\emph{a}} & \ce{BiOCl}, \ce{NaTaOCl4} \\
Halide                    & Shannon   & 0.58 & \ce{Li2ZrCl6}, \ce{CaF2}, \ce{AlF3} \\
Large-cation nitride      & Shannon   & 0.52 & \ce{InN}, \ce{ZrN}, \ce{HfN}, \ce{ScN} \\
Small-cation nitride      & Shannon   & 0.62 & \ce{AlN}, \ce{GaN}, \ce{TiN}, \ce{BN}, \ce{Si3N4} \\
Hydride                   & Shannon   & 0.55 & \ce{LiH}, \ce{MgH2} \\
Chalcogenide              & Cordero   & 0.30 & \ce{ZnS}, \ce{CdTe}, \ce{GeTe}, \ce{Sb2Te3} \\
Chalcogenide glass        & Cordero   & 0.23 & \ce{GeS2}, \ce{GeSe2}, \ce{As2S3}, \ce{As2Se3} \\
Pnictide                  & Cordero   & 0.32 & \ce{GaAs}, \ce{InP}, \ce{GaSb} \\
Tetrahedral element       & Cordero   & 0.30 & \ce{Si}, \ce{Ge} \\
Low-coordinated element   & Cordero   & 0.28 & \ce{C}, \ce{P}, \ce{As}, \ce{Se} \\
Hydrogenated network      & Cordero   & 0.28--0.32\textsuperscript{\emph{b}} & a-Si:H, a-C:H \\
Covalent carbide          & Cordero   & 0.32 & \ce{SiC}, \ce{B4C} \\
Transition-metal carbide  & Goldschmidt; Cordero C & 0.60 & \ce{TiC}, \ce{WC} \\
Boride                    & Goldschmidt; Cordero B & 0.60 & \ce{TiB2}, \ce{MgB2} \\
Alloy                     & Goldschmidt & 0.60 & \ce{NiTi}, \ce{CuZr} \\
Default                   & Shannon   & 0.52 & compositions outside the classes above \\
\bottomrule
\end{tabular}

  \smallskip\noindent\footnotesize \textsuperscript{\emph{a}} Interpolated between the factor of the halogen-free composition and 0.58 by the halogen fraction of the anions.
  \par\noindent\textsuperscript{\emph{b}} The factor of the H-free host (0.30 a-Si:H, 0.28 a-C:H, 0.32 a-SiC:H), with H at \SI{0.90}{\angstrom}.
\end{table}

\section{Density estimation and calibration}
\label{sec:packing-calibration}

\begin{figure}[H]
\centering
\includegraphics[width=0.55\textwidth]{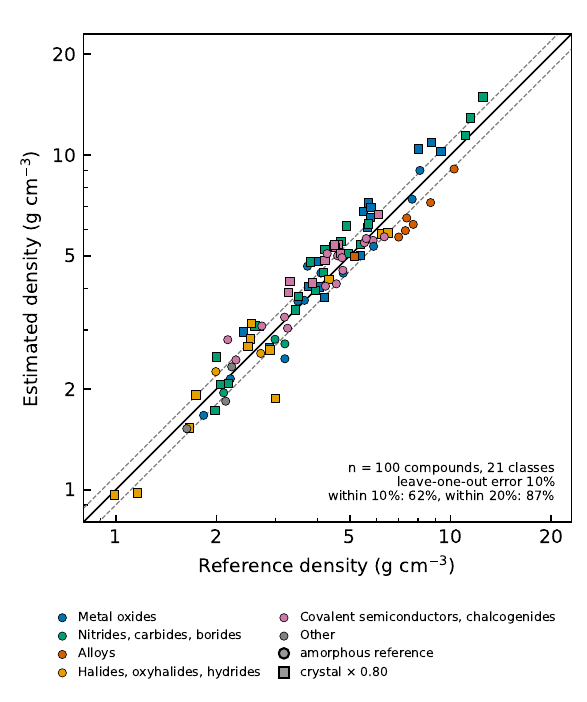}
\caption{Sphere-packing density estimate against the target densities of
the 100 calibration compounds (circles: amorphous references; squares:
crystal density $\times$ 0.80), colored by class group; dashed lines,
\SI{\pm 10}{\percent}. Data and per-class fit in Tables~\ref{tab:reference-densities} and
\ref{tab:packing-fit}.}
\label{fig:parity}
\end{figure}

The density estimate uses Eq.~\mainref{eq:cell}{(1)} of the main text,
with the radii and packing factor $f_{\mathrm{pack}}$ of the material
class (Table~\ref{tab:packing}). We compared the shipped factors with
the 100 target densities of Table~\ref{tab:reference-densities}:
41 measured on amorphous samples and 59 assumed to be 0.80 times the crystal
density. The error therefore measures agreement with this mixed set,
not with 100 amorphous measurements. Each compound gives a back-solved
factor $f_i = f_{\mathrm{code}}\,\rho_{\mathrm{target},i}/\rho_{\mathrm{est},i}$,
and the fitted class factor is the median of $f_i$ over the class
(Table~\ref{tab:packing-fit}). The shipped factors give a mean absolute
relative error, the mean of $|\rho_{\mathrm{est}}/\rho_{\mathrm{target}}-1|$,
of \SI{11.9}{\percent} (Figure~\ref{fig:parity}). For the fitted
factors, leave-one-out cross-validation, which predicts each compound
from the median of the other members of its class, gives
\SI{9.8}{\percent} over 99 compounds. \ce{BeO} is excluded as the only
small-cation oxide.

The script \path{examples/calibration/fit_packing_factors.py} and its
input \path{examples/calibration/packing_reference_densities.csv} in
the software repository reproduce these results with AmorphGen v1.0.0.
The fitted factors are reported for comparison and do not replace the
shipped ones. Where the two differ most, the estimate is systematically
off: metallic glasses start about \SI{18}{\percent} below their
measured densities (0.60 against a fitted 0.73), and fluorite-type
dioxides about \SI{24}{\percent} above (0.63 against 0.51). For such
systems, a measured density should be supplied, or the fitted factor
used.

\newpage

\subsection{Reference densities}
\label{sec:reference-densities}

The rows follow the material-class order of Table~\ref{tab:packing}.
All density columns use \si{\gram\per\cubic\centi\metre}.

\begingroup
\footnotesize
\setlength{\tabcolsep}{5pt}
\begin{longtable}{@{}l l l r r r l@{}}
\caption{The 100 sourced reference densities of the calibration set. Kind: amorphous (A) or crystalline (C) reference; target = reference for A, reference $\times$ 0.80 for C; estimate = AmorphGen's class estimate for the composition (\si{\gram\per\cubic\centi\metre}). Handbook values are taken from the CRC Handbook, 97th edition \cite{haynes2016crc}.}\label{tab:reference-densities}\\
\toprule
Compound & Class & Kind & Reference & Target & Estimate & Source \\
\midrule
\endfirsthead
\toprule
Compound & Class & Kind & Reference & Target & Estimate & Source \\
\midrule
\endhead
\bottomrule
\endfoot

\ce{As2O3} & network-forming oxide & A & 3.74 & 3.74 & 4.65 & \cite{pubchem_as2o3} \\
\ce{B2O3} & network-forming oxide & A & 1.83 & 1.83 & 1.67 & \cite{napolitano1965} \\
\ce{GeO2} & network-forming oxide & A & 3.66 & 3.66 & 3.68 & \cite{micoulaut2006} \\
\ce{SiO2} & network-forming oxide & A & 2.20 & 2.20 & 2.15 & \cite{bruckner1970} \\
\ce{TeO2} & network-forming oxide & A & 5.65 & 5.65 & 5.59 & \cite{alderman2020teo2} \\
\ce{Al2O3} & metal oxide & A & 3.20 & 3.20 & 2.46 & \cite{pugliese2022} \\
\ce{Fe2O3} & metal oxide & C & 5.25 & 4.20 & 3.75 & \cite{haynes2016crc} \\
\ce{Ga2O3} & metal oxide & A & 4.78 & 4.78 & 4.44 & \cite{liu2023} \\
\ce{InGaZnO4} & metal oxide & A & 5.89 & 5.89 & 5.33 & \cite{kumara2021} \\
\ce{MgO} & metal oxide & C & 3.60 & 2.88 & 2.67 & \cite{haynes2016crc} \\
\ce{NiO} & metal oxide & C & 6.72 & 5.38 & 5.01 & \cite{haynes2016crc} \\
\ce{Sm2O3} & metal oxide & C & 7.09 & 5.67 & 7.20 & \cite{zavyalova1976} \\
\ce{Y2O3} & metal oxide & C & 5.03 & 4.02 & 4.80 & \cite{haynes2016crc} \\
\ce{ZnO} & metal oxide & C & 5.60 & 4.48 & 5.33 & \cite{haynes2016crc} \\
\ce{IrO2} & rutile-type dioxide & C & 11.70 & 9.36 & 10.23 & \cite{haynes2016crc} \\
\ce{MnO2} & rutile-type dioxide & C & 5.08 & 4.06 & 4.04 & \cite{haynes2016crc} \\
\ce{RuO2} & rutile-type dioxide & C & 7.05 & 5.64 & 6.08 & \cite{haynes2016crc} \\
\ce{SnO2} & rutile-type dioxide & C & 6.85 & 5.48 & 6.78 & \cite{haynes2016crc} \\
\ce{TiO2} & rutile-type dioxide & A & 3.50 & 3.50 & 3.66 & \cite{mergel2000} \\
\ce{CeO2} & fluorite-type dioxide & C & 7.22 & 5.77 & 6.99 & \cite{haynes2016crc} \\
\ce{HfO2} & fluorite-type dioxide & A & 8.10 & 8.10 & 8.99 & \cite{kim2020hfo2} \\
\ce{ThO2} & fluorite-type dioxide & C & 10.00 & 8.00 & 10.44 & \cite{haynes2016crc} \\
\ce{UO2} & fluorite-type dioxide & C & 10.97 & 8.78 & 10.89 & \cite{haynes2016crc} \\
\ce{MoO3} & high-valent oxide & C & 4.70 & 3.76 & 4.06 & \cite{haynes2016crc} \\
\ce{Nb2O5} & high-valent oxide & A & 4.10 & 4.10 & 4.44 & \cite{shank2018} \\
\ce{Ta2O5} & high-valent oxide & A & 7.68 & 7.68 & 7.38 & \cite{alderman2018tantala} \\
\ce{V2O5} & high-valent oxide & C & 3.35 & 2.68 & 3.08 & \cite{haynes2016crc} \\
\ce{WO3} & high-valent oxide & C & 7.20 & 5.76 & 6.53 & \cite{haynes2016crc} \\
\ce{BeO} & small-cation oxide & C & 3.01 & 2.41 & 2.96 & \cite{haynes2016crc} \\
\ce{BiOBr} & oxyhalide & C & 8.12 & 6.50 & 5.85 & \cite{ketterer1986} \\
\ce{BiOCl} & oxyhalide & C & 7.78 & 6.22 & 5.81 & \cite{handbook_mineralogy_bismoclite} \\
\ce{FeOCl} & oxyhalide & C & 3.61 & 2.89 & 2.62 & \cite{lind1970} \\
\ce{LaOCl} & oxyhalide & C & 5.41 & 4.33 & 4.25 & \cite{maslen1996} \\
\ce{AlF3} & halide & C & 3.10 & 2.48 & 2.68 & \cite{haynes2016crc} \\
\ce{BeF2} & halide & A & 1.99 & 1.99 & 2.25 & \cite{wright1989} \\
\ce{CaF2} & halide & C & 3.18 & 2.54 & 3.15 & \cite{haynes2016crc} \\
\ce{LiCl} & halide & C & 2.07 & 1.66 & 1.53 & \cite{haynes2016crc} \\
\ce{MgF2} & halide & C & 3.15 & 2.52 & 2.82 & \cite{haynes2016crc} \\
\ce{NaCl} & halide & C & 2.17 & 1.74 & 1.92 & \cite{haynes2016crc} \\
\ce{ZnCl2} & halide & A & 2.71 & 2.71 & 2.55 & \cite{desa1982} \\
\ce{HfN} & large-cation nitride & C & 13.81 & 11.05 & 11.44 & \cite{seo2004} \\
\ce{ScN} & large-cation nitride & C & 4.30 & 3.44 & 3.45 & \cite{gall2001} \\
\ce{YN} & large-cation nitride & C & 5.88 & 4.70 & 5.52 & \cite{du2010} \\
\ce{ZrN} & large-cation nitride & C & 7.09 & 5.67 & 6.22 & \cite{haynes2016crc} \\
\ce{AlN} & small-cation nitride & C & 3.25 & 2.60 & 3.09 & \cite{haynes2016crc} \\
\ce{BN} & small-cation nitride & A & 2.10 & 2.10 & 1.95 & \cite{sattari2025} \\
\ce{Be3N2} & small-cation nitride & C & 2.71 & 2.17 & 2.08 & \cite{haynes2016crc} \\
\ce{GaN} & small-cation nitride & C & 6.10 & 4.88 & 6.14 & \cite{haynes2016crc} \\
\ce{Ge3N4} & small-cation nitride & C & 5.26 & 4.21 & 5.22 & \cite{ruddlesden1958} \\
\ce{Si3N4} & small-cation nitride & A & 3.20 & 3.20 & 2.73 & \cite{yen2003} \\
\ce{TiN} & small-cation nitride & C & 5.21 & 4.17 & 4.46 & \cite{haynes2016crc} \\
\ce{LiH} & hydride & C & 0.78 & 0.62 & 0.54 & \cite{haynes2016crc} \\
\ce{MgH2} & hydride & C & 1.45 & 1.16 & 0.98 & \cite{haynes2016crc} \\
\ce{NaAlH4} & hydride & C & 1.24 & 0.99 & 0.97 & \cite{haynes2016crc} \\
\ce{TiH2} & hydride & C & 3.75 & 3.00 & 1.88 & \cite{haynes2016crc} \\
\ce{As40Te60} & chalcogenide & A & 5.60 & 5.60 & 5.63 & \cite{cornet1973} \\
\ce{CdS} & chalcogenide & C & 4.83 & 3.86 & 4.15 & \cite{haynes2016crc} \\
\ce{CdTe} & chalcogenide & C & 5.85 & 4.68 & 5.08 & \cite{shah2021} \\
\ce{Ge2Sb2Te5} & chalcogenide & A & 5.87 & 5.87 & 5.56 & \cite{njoroge2002} \\
\ce{GeTe} & chalcogenide & A & 5.53 & 5.53 & 5.47 & \cite{ghezzi2011} \\
\ce{PbS} & chalcogenide & C & 7.60 & 6.08 & 6.66 & \cite{haynes2016crc} \\
\ce{ZnS} & chalcogenide & C & 4.09 & 3.27 & 3.90 & \cite{haynes2016crc} \\
\ce{ZnSe} & chalcogenide & C & 5.27 & 4.22 & 4.84 & \cite{madelung2004} \\
\ce{As2S3} & chalcogenide glass & A & 3.19 & 3.19 & 3.28 & \cite{knotek2020} \\
\ce{As2Se3} & chalcogenide glass & A & 4.56 & 4.56 & 4.12 & \cite{knotek2020} \\
\ce{GeS2} & chalcogenide glass & A & 2.73 & 2.73 & 3.08 & \cite{tverjanovich2021} \\
\ce{GeSe2} & chalcogenide glass & A & 4.23 & 4.23 & 4.06 & \cite{tverjanovich2021} \\
\ce{Ga9Sb91} & pnictide & A & 6.34 & 6.34 & 5.70 & \cite{putero2014} \\
\ce{GaP} & pnictide & C & 4.14 & 3.31 & 4.20 & \cite{haynes2016crc} \\
\ce{GaSb} & pnictide & C & 5.61 & 4.49 & 5.40 & \cite{haynes2016crc} \\
\ce{InAs} & pnictide & C & 5.67 & 4.54 & 5.29 & \cite{haynes2016crc} \\
\ce{InP} & pnictide & A & 4.77 & 4.77 & 4.52 & \cite{cliche1994} \\
\ce{InSb} & pnictide & C & 5.77 & 4.62 & 5.41 & \cite{haynes2016crc} \\
\ce{Ge} & tetrahedral element & A & 4.60 & 4.60 & 5.00 & \cite{donovan1970} \\
\ce{Si} & tetrahedral element & A & 2.29 & 2.29 & 2.44 & \cite{custer1994} \\
\ce{As} & low-coordinated element & A & 4.74 & 4.74 & 4.93 & \cite{smith1975} \\
\ce{C} & low-coordinated element & A & 3.26 & 3.26 & 3.04 & \cite{libassi2000} \\
\ce{P} & low-coordinated element & A & 2.16 & 2.16 & 2.81 & \cite{haynes2016crc} \\
\ce{Se} & low-coordinated element & A & 4.28 & 4.28 & 5.07 & \cite{haynes2016crc} \\
\ce{C60H40} & hydrogenated network & A & 1.63 & 1.63 & 1.52 & \cite{libassi2000} \\
\ce{C70H30} & hydrogenated network & A & 2.13 & 2.13 & 1.84 & \cite{libassi2000} \\
\ce{Si91H9} & hydrogenated network & A & 2.22 & 2.22 & 2.33 & \cite{remes1998} \\
\ce{B4C} & covalent carbide & C & 2.50 & 2.00 & 2.49 & \cite{haynes2016crc} \\
\ce{Be2C} & covalent carbide & C & 2.47 & 1.98 & 1.73 & \cite{campbell2023} \\
\ce{SiC} & covalent carbide & A & 2.99 & 2.99 & 2.82 & \cite{heera1997} \\
\ce{TaC} & transition-metal carbide & C & 14.30 & 11.44 & 12.92 & \cite{haynes2016crc} \\
\ce{TiC} & transition-metal carbide & C & 4.93 & 3.94 & 3.94 & \cite{haynes2016crc} \\
\ce{WC} & transition-metal carbide & C & 15.60 & 12.48 & 14.91 & \cite{haynes2016crc} \\
\ce{ZrC} & transition-metal carbide & C & 6.73 & 5.38 & 5.41 & \cite{haynes2016crc} \\
\ce{LaB6} & boride & C & 4.76 & 3.81 & 4.80 & \cite{haynes2016crc} \\
\ce{MgB2} & boride & C & 2.57 & 2.06 & 2.07 & \cite{haynes2016crc} \\
\ce{TiB2} & boride & C & 4.38 & 3.50 & 3.79 & \cite{haynes2016crc} \\
\ce{ZrB2} & boride & C & 6.17 & 4.94 & 5.08 & \cite{haynes2016crc} \\
\ce{Cu64Zr36} & alloy & A & 7.74 & 7.74 & 6.21 & \cite{wong2018} \\
\ce{CuZr} & alloy & A & 7.33 & 7.33 & 5.94 & \cite{wong2018} \\
\ce{CuZr2} & alloy & A & 7.00 & 7.00 & 5.69 & \cite{wong2018} \\
\ce{Fe80B20} & alloy & A & 7.40 & 7.40 & 6.48 & \cite{hasegawa1978} \\
\ce{Ni60Nb40} & alloy & A & 8.72 & 8.72 & 7.21 & \cite{basak1979} \\
\ce{NiTi} & alloy & C & 6.46 & 5.17 & 4.99 & \cite{otsuka2005} \\
\ce{Pd4Si} & alloy & A & 10.25 & 10.25 & 9.08 & \cite{polk1972} \\
\end{longtable}
\endgroup

\subsection{Fitted packing factors by class}
\label{sec:packing-classes}

The fitted factors and errors are grouped in the same class order as
Table~\ref{tab:packing}. MAE below denotes the mean absolute relative
error defined above, expressed as a percentage. No leave-one-out
error is reported for the class containing only \ce{BeO}.

\begingroup
\footnotesize
\setlength{\tabcolsep}{5pt}
\begin{longtable}{@{}l r r r r r r@{}}
\caption{Per-class fit results: the number of compounds ($n$, including $n_\mathrm{A}$ amorphous references), the packing factor in the code, the median back-solved factor, its range over the class, and the mean absolute error (MAE) of the class median in leave-one-out cross-validation. For the oxyhalides, $f_{\mathrm{code}}$ is interpolated by the halogen fraction of the anions (Table~\ref{tab:packing}) and is 0.55 for all four compounds; for the hydrogenated networks, it is the factor of the H-free host (0.28 for a-C:H, 0.30 for a-Si:H).}\label{tab:packing-fit}\\
\toprule
Class & $n$ & $n_\mathrm{A}$ & $f_{\mathrm{code}}$ & $f_{\mathrm{fit}}$ (median) & range & MAE \\
\midrule
\endfirsthead
\toprule
Class & $n$ & $n_\mathrm{A}$ & $f_{\mathrm{code}}$ & $f_{\mathrm{fit}}$ (median) & range & MAE \\
\midrule
\endhead
\bottomrule
\endfoot

network-forming oxide & 5 & 5 & 0.50 & 0.51 & 0.40--0.55 & \SI{8.0}{\percent} \\
metal oxide & 9 & 3 & 0.52 & 0.56 & 0.41--0.68 & \SI{13.2}{\percent} \\
rutile-type dioxide & 5 & 1 & 0.66 & 0.61 & 0.53--0.66 & \SI{6.5}{\percent} \\
fluorite-type dioxide & 4 & 1 & 0.63 & 0.51 & 0.48--0.57 & \SI{5.8}{\percent} \\
high-valent oxide & 5 & 2 & 0.60 & 0.55 & 0.52--0.62 & \SI{5.8}{\percent} \\
small-cation oxide & 1 & 0 & 0.35 & 0.28 & 0.28--0.28 & -- \\
oxyhalide & 4 & 0 & 0.55 & 0.60 & 0.56--0.61 & \SI{4.5}{\percent} \\
halide & 7 & 2 & 0.58 & 0.52 & 0.47--0.63 & \SI{7.9}{\percent} \\
large-cation nitride & 4 & 0 & 0.52 & 0.49 & 0.44--0.52 & \SI{8.4}{\percent} \\
small-cation nitride & 7 & 2 & 0.62 & 0.58 & 0.49--0.73 & \SI{17.3}{\percent} \\
hydride & 4 & 0 & 0.55 & 0.64 & 0.56--0.88 & \SI{12.6}{\percent} \\
chalcogenide & 8 & 3 & 0.30 & 0.28 & 0.25--0.32 & \SI{6.4}{\percent} \\
chalcogenide glass & 4 & 4 & 0.23 & 0.23 & 0.20--0.25 & \SI{10.9}{\percent} \\
pnictide & 6 & 2 & 0.32 & 0.27 & 0.25--0.36 & \SI{9.1}{\percent} \\
tetrahedral element & 2 & 2 & 0.30 & 0.28 & 0.28--0.28 & \SI{1.7}{\percent} \\
low-coordinated element & 4 & 4 & 0.28 & 0.25 & 0.22--0.30 & \SI{18.0}{\percent} \\
hydrogenated network & 3 & 3 & 0.28, 0.30 & 0.30 & 0.29--0.32 & \SI{6.9}{\percent} \\
covalent carbide & 3 & 1 & 0.32 & 0.34 & 0.26--0.37 & \SI{21.5}{\percent} \\
transition-metal carbide & 4 & 0 & 0.60 & 0.56 & 0.50--0.60 & \SI{13.4}{\percent} \\
boride & 4 & 0 & 0.60 & 0.57 & 0.48--0.60 & \SI{9.8}{\percent} \\
alloy & 7 & 6 & 0.60 & 0.73 & 0.62--0.75 & \SI{7.1}{\percent} \\
\end{longtable}
\endgroup

\section{Structural ensembles and simulation settings}
\label{sec:provenance}

This section records the settings of the ensembles used for the
structural comparisons in \mainref{examples}{Section~3} of the main text.
Table~\ref{tab:ensemble-provenance} gives the engines and task splits;
the regenerated a-\ce{Ga2O3} and separate a-\ce{InGaZnO4} ensembles
used for EXAFS are described in Sections~\ref{sec:exafs-ga2o3} and \ref{sec:exafs-igzo}.

\paragraph{Default protocol and example overrides.}
The seven stages of the melt-quench pipeline are optimization,
low-temperature equilibration, heating, liquid equilibration,
quenching, low-temperature equilibration, and final optimization.
The default low and melt temperatures are \SIlist{300;3000}{\kelvin},
with heating and cooling at \SI{100}{\kelvin\per\pico\second}, a
\SI{0.5}{\femto\second} timestep, and a
\SI{10}{\pico\second} liquid hold. Heating and liquid equilibration
use a barostat by default.

The hybrid example in \mainref{sec:code}{Section~2.3} of the main
text uses fixed-volume, single-precision MD:
\SI{50}{\pico\second} at \SI{3000}{\kelvin}, cooling to
\SI{300}{\kelvin} at \SI{100}{\kelvin\per\pico\second}, and
\SI{10}{\pico\second} at \SI{300}{\kelvin}, followed by final
relaxation. The melt-quench example heats to \SI{3000}{\kelvin}
and holds the liquid for \SI{50}{\pico\second}, both at fixed
volume, then quenches up to 20 decorrelated snapshots and
equilibrates them under NPT at \SI{300}{\kelvin} before relaxation.
The analysis command checks every descriptor with a range in the
reference file: for a-\ce{Ga2O3}, density, Ga--O bond length,
coordination numbers, and bond angles.

\paragraph{Divergence guards.}
Any MD stage or relaxation stops with a message if a position,
energy, or force becomes non-finite, two atoms come within
\SI{0.5}{\angstrom}, the energy changes by more than
\SI{10}{\electronvolt} per atom between checks, the temperature
exceeds \SI{e5}{\kelvin}, or the volume falls below one fifth or
rises above five times its starting value.

\paragraph{Integration and numerical precision.}
The MD used a Langevin thermostat with a friction of
\SI{0.01}{\per\femto\second} and a timestep of
\SI{0.5}{\femto\second} (\SI{1}{\femto\second} for a-As).
For the MACE calculations, random-ensemble relaxations used double
precision; hybrid MD, final hybrid relaxations, and subsequent
cubic-cell relaxations used single precision. The BKS calculations
used double precision: placement, fixed-cell relaxation, and MD ran
through ASE \cite{larsen2017} on the CPU, and subsequent cubic-cell relaxation used
TorchSim \cite{torchsim2025} on the CPU. The BKS pair parameters, Ewald settings, and
short-range repulsive core are specified in
\path{examples/softwarex_bks_sio2.yaml} of the software repository;
the cell-relaxation command is in
\path{examples/softwarex_run_sio2_bks_cubic.slurm}.

\paragraph{Silica and arsenic preparation.}
The silica comparison used the same initial placements for BKS and
MACE-MPA-0. BKS used Ewald summation \cite{ewald1921} with a
\SI{10}{\angstrom} cutoff; its fixed-cell relaxation and anneal ran
at the generation density of \SI{2.20}{\gram\per\cubic\centi\metre}.
The MACE-MPA-0 anneal used the relaxed density of
\SI{2.12}{\gram\per\cubic\centi\metre}. The densities in
\mainref{tab:glasses}{Table~2} of the main text are after separate
cubic-cell relaxations. Potential and annealing density therefore
both differ in this comparison.

The five 2000-atom a-As cells were placed at the measured density of
\SI{4.74}{\gram\per\cubic\centi\metre} \cite{smith1975}, with a
coordination target of three (\texttt{-{}-target-cn As=3}, also the
automatic target in v1.0.0).
The hybrid route annealed them for \SI{50}{\pico\second} at
\SI{1500}{\kelvin} with a \SI{1}{\femto\second} timestep and quenched
them at \SI{100}{\kelvin\per\pico\second}, followed by
\SI{10}{\pico\second} at \SI{300}{\kelvin}. The whole
\SI{72}{\pico\second} protocol is about 20 times shorter than that
of Liu \textit{et al.}\ \cite{liu2026as}, whose 2000-atom cells were
placed with \texttt{buildcell} at \SI{4.74}{\gram\per\cubic\centi\metre}
and then run for about \SI{1.4}{\nano\second} of melt-quench dynamics
with their r$^2$SCAN-trained MACE potential. They analyzed one such
model, deposited with their paper, and estimated uncertainties from
ten independent runs. Our reference values are computed from the
deposited structure.

\paragraph{Seeds and task allocation.}
The random ensembles were generated from global seed 2026. The oxide
hybrids used the first 20 relaxed structures (indices 0--19). Each
task $T$ used seed $2026 + T$: tasks 1 and 2 took ten structures each
for a-\ce{Ga2O3}, a-\ce{TiO2}, and a-\ce{InGaZnO4}; tasks 1--4 took
five each for a-\ce{SiO2} with MACE-MPA-0; and tasks 4--23 took one
each for BKS. These assignments are implemented in
\path{examples/softwarex_run_oxide.slurm} and
\path{examples/softwarex_run_sio2_bks.slurm}. The a-As hybrids
used seed 2026. The a-As MACE-MPA-0 hybrids started from the same
placements as the system-specific MACE calculations, relaxed again
with MACE-MPA-0.

For the BKS random ensemble, tasks 0--3 generated indices 0--24,
25--49, 50--74, and 75--99, respectively, using the same global seed.
The task-dependent seeds above apply to the subsequent hybrid runs.

\begin{table}[H]
\centering
\caption{Engine and task split of the structural ensembles. Each entry
gives tasks $\times$ structures per task; $4+1$ denotes two jobs,
one with four structures and one with a single structure. The BKS GPU timing comparisons in
Table~\ref{tab:timings} are separate from the structural ensembles listed here.}
\label{tab:ensemble-provenance}
\footnotesize
\setlength{\tabcolsep}{4pt}
\begin{tabular}{@{}l l l l@{}}
\toprule
System & Potential & Ensemble & Engine; tasks $\times$ structures \\
\midrule
a-\ce{Ga2O3} & MACE-MPA-0 & random + relax & TorchSim GPU; $1\times100$ \\
 & & hybrid & TorchSim GPU; $2\times10$ \\
a-\ce{TiO2} & MACE-MPA-0 & random + relax & TorchSim GPU; $1\times100$ \\
 & & hybrid & TorchSim GPU; $2\times10$ \\
a-\ce{InGaZnO4} & MACE-MPA-0 & random + relax & TorchSim GPU; $1\times100$ \\
 & & hybrid & TorchSim GPU; $2\times10$ \\
a-\ce{SiO2} & MACE-MPA-0 & random + relax & TorchSim GPU; $1\times100$ \\
 & & hybrid & TorchSim GPU; $4\times5$ \\
a-\ce{SiO2} & BKS & random + relax & ASE CPU; $4\times25$ \\
 & & hybrid & ASE CPU; $20\times1$ \\
a-As & MACE (Liu \textit{et al.}) & random + relax, hybrid & TorchSim GPU; $1\times5$ \\
a-As & MACE-MPA-0 & random + relax & TorchSim GPU; $1\times5$ \\
 & & hybrid & TorchSim GPU; $4+1$ \\
a-As (all four) & both & cubic-cell relaxation & TorchSim GPU; $4\times5$ \\
a-\ce{SiO2} & BKS & cubic-cell relaxation & TorchSim CPU \\
\bottomrule
\end{tabular}
\end{table}

\section{Structural analysis and additional comparisons}
\label{sec:analysis}

\paragraph{Neighbor cutoffs and dimers.}
Neighbor cutoffs are placed at the first minimum of each partial
$g(r)$ of the ensemble: Ga--O \SI{2.47}{\angstrom} in a-\ce{Ga2O3},
Ti--O \SI{2.61}{\angstrom}, In--O, Ga--O, and Zn--O
\SIlist{2.73;2.49;2.55}{\angstrom} in a-\ce{InGaZnO4}, and Si--O
\SI{2.24}{\angstrom}. These oxide values are for the MACE-MPA-0 hybrid
ensembles. Cutoffs vary between ensembles: In--O spans
\SIrange{2.73}{2.81}{\angstrom} and Si--O spans
\SIrange{2.13}{2.24}{\angstrom}, including BKS. For a-As, the
\SI{2.9}{\angstrom} cutoff of Liu \textit{et al.}\ \cite{liu2026as}
is used. Coordination numbers, angles, and rings use these cutoffs.
An O--O pair closer than 0.85 times its minimum placement separation
is counted as a dimer. For the oxides here, the O--O minimum separation
is \SI{2.24}{\angstrom}, so the threshold is \SI{1.9}{\angstrom}.

\paragraph{Ring statistics.}
Each primitive ring (a ring with no shortcut through the network) is
counted once, with cations as nodes bridged by O (As atoms bonded to
As for a-As). Ring sizes are limited to 8 members for a-\ce{Ga2O3},
a-\ce{TiO2}, and a-\ce{InGaZnO4}, 12 for a-\ce{SiO2}, and 10 for a-As.
Counts are reported per node. These are the settings of the figure
scripts; by default, \texttt{-{}-rings} assigns each network edge its
shortest ring, up to 12 members. In the oxides
other than a-\ce{SiO2}, an O bonded to three cations closes a
three-membered ring, so their ring counts mainly reflect the O
coordination.

\paragraph{Structure factors.}
$S(q)$ is computed directly on the reciprocal-lattice
vectors of each cell with Faber--Ziman \cite{faber1965} X-ray weighting and
$q$-dependent Waasmaier--Kirfel form factors \cite{waasmaier1995}, in \SI{0.04}{\per\angstrom}
bins smoothed with a Gaussian of \SI{0.08}{\per\angstrom} and averaged over 20
structures per ensemble (all five for a-As). Bins below
$1.3 \times 2\pi/L$ are dropped, which for the 648-atom a-\ce{SiO2}
cells ($L = \SI{21.4}{\angstrom}$) leaves
$q \ge \SI{0.38}{\per\angstrom}$. These are the settings of the figure
scripts; the defaults of \texttt{amorphgen -{}-analyse}
(\SI{0.05}{\per\angstrom} bins and smoothing, no low-$q$ cut) give
slightly different curves. With the \SI{0.10}{\per\angstrom} smoothing
of the cluster analyzes, the a-\ce{SiO2} FSDP heights change by at most
0.02 and the positions by at most \SI{0.08}{\per\angstrom}.

\paragraph{Statistical uncertainty.}
Each independent structure contributes one mean value for a descriptor,
and these structure means receive equal weight. The standard error is
$\mathrm{SE} = s/\sqrt{n}$, where $s$ is the sample standard deviation
of the structure means and $n$ is the number of contributing structures.
The \SI{95}{\percent} Student-$t$ confidence interval is
$\bar{x} \pm t_{0.975,n-1}\,\mathrm{SE}$, with full width
$W_{95} = 2t_{0.975,n-1}\,\mathrm{SE}$.
The reported full interval widths are below \SI{0.004}{\angstrom}
for mean bond lengths, \ang{1} for mean angles, and 0.05 for mean
coordination numbers, for both the five-cell a-As ensembles and the
oxide ensembles. These intervals quantify uncertainty in the ensemble mean;
the spread of individual bond lengths, angles, or site coordinations
is a separate quantity. The bootstrap intervals used for EXAFS are
specified in Section~\ref{sec:exafs}.

\paragraph{Displacements during annealing.}
Over the \SI{50}{\pico\second} anneal, the mean unwrapped displacement of the atoms is
\SIrange{12}{13}{\angstrom} in a-\ce{Ga2O3}, a-\ce{TiO2}, and a-\ce{InGaZnO4},
\SIrange{10}{11}{\angstrom} in a-As at \SI{1500}{\kelvin}, and only \SIrange{2.4}{3.5}{\angstrom} in
a-\ce{SiO2} with either potential.

\subsection{Additional structural comparisons}
\label{sec:structural-comparisons}

The numerical descriptors are collected in
\mainref{tab:oxides}{Tables~1} and \mainref{tab:glasses}{2} of the main
text. The comparisons below provide reference provenance and further
details of \mainref{fig:grid}{Figures~3} and
\mainref{fig:anneal}{4}.

\paragraph{Metal oxides.}
The 20 a-\ce{Ga2O3} reference structures of
Ref.~\cite{kaewmeechai2025} were made by melt-quench with the
classical potential of Blanco \textit{et al.}\
\cite{blanco2005energetics} and optimized with PBE0-TC-LRC.
Both AmorphGen ensembles reproduce their Ga coordination and split
between four- and five-fold sites, but have Ga--O bonds about
\SI{0.03}{\angstrom} longer. The X-ray $S(q)$ of both ensembles matches the peak
positions of the reference set and the measured film \cite{liu2023}
from \SI{2.5}{\per\angstrom} upward, with a first peak
\SIrange{15}{20}{\percent} lower. For a-\ce{TiO2}, the first peak is
at 1.95 against \SI{2.28}{\per\angstrom} measured \cite{petkov1998}.

The a-\ce{InGaZnO4} EXAFS reference coordination numbers are
\num{5.1 \pm 0.7} (In), \num{3.9 \pm 0.6} (Ga), and
\num{3.7 \pm 0.5} (Zn) \cite{kumara2021}.
The calculated metal--O distances are
\SIrange{0.06}{0.08}{\angstrom} longer than measured, in the direction
expected from the PBE data on which MACE-MPA-0 is trained
\cite{haas2009lattice}; annealing does not remove this difference.
The sphere-packing estimates would start a-\ce{Ga2O3} and
a-\ce{InGaZnO4} at \SIlist{4.44;5.33}{\gram\per\cubic\centi\metre},
\SIrange{7}{13}{\percent} below the measured densities, and
a-\ce{TiO2} slightly above its film density range
(Table~\ref{tab:reference-densities}).

\paragraph{Silica.}
With MACE-MPA-0, every random + relax structure contains O--O dimers;
annealing removes them from 16 of the 20 cells. The low-$q$ rise
before annealing indicates density fluctuations. BKS silica
simulations in the literature cool over nanoseconds
\cite{vollmayr1996}; we did not test whether the shorter protocol
used here limits the FSDP, whose origin remains debated
\cite{elliott1991fsdp}.

\paragraph{Arsenic.}
With the potential of Liu \textit{et al.}\ \cite{liu2026as},
relaxation alone reproduces their model's $S(q)$ above
\SI{2}{\per\angstrom}. MACE-MPA-0 gives an As--As bond
\SI{0.04}{\angstrom} longer and twice as many four-fold sites
(16 against \SI{8}{\percent}). With the system-specific potential,
annealing removes half the four-fold sites in excess of the reference
and raises the number of five-membered rings from 0.14 to 0.17 per
atom, against 0.22 in the reference. With MACE-MPA-0,
\SI{14}{\percent} of As remains four-fold after annealing.
After cubic-cell relaxation, all four ensembles have densities
\SIrange{3}{7}{\percent} below experiment and within
\SI{0.14}{\gram\per\cubic\centi\metre} of the reference model.
Across the five cells annealed with the system-specific potential,
FSDP heights range from 0.38 to 0.44, below both the deposited reference
model (0.51 as computed here; \num{0.56 \pm 0.01} over ten runs in
Ref.~\cite{liu2026as}) and the measured value (0.55) of
Ref.~\cite{smith1975}, from combined X-ray and neutron data as
digitized in Ref.~\cite{liu2026as}.

\section{Computational cost of structure generation}
\label{sec:cost}

Table~\ref{tab:timings} lists the wall times of the calculations in
\mainref{examples}{Section~3} of the main text, taken from the job logs.
The ensemble definitions and task splits are given in
Section~\ref{sec:provenance}. GPU jobs ran on
one NVIDIA A100-SXM4 GPU with four Intel Xeon Ice Lake cores. BKS jobs
on the CPU ran on four Intel Xeon Platinum cores; the BKS structures
analyzed in the main text come from these runs. The BKS rows marked
GPU repeat the same protocols with the batched engine (Ewald summation)
for comparison and are not used in the main text.

\begin{table}[H]
\centering
\caption{Wall time by system, potential, and route. Brackets in the
Structures column give the number of jobs $\times$ structures per job;
two wall times or a range are those of the separate jobs. Per-structure
times are the summed job wall times divided by the number of structures,
rather than the elapsed time of concurrently running jobs. For
random + relax, the wall time includes the placement on the CPU, and
the time in parentheses is the relaxation on the GPU.}
\label{tab:timings}
\footnotesize
\resizebox{\textwidth}{!}{%
\begin{tabular}{@{}l r l l l l r r@{}}
\toprule
System & Atoms & Potential & Route & Structures & Hardware & Wall time & Per structure \\
\midrule
a-\ce{Ga2O3} & 400 & MACE-MPA-0 & random + relax & 100 & GPU & \SI{0.8}{\hour} (\SI{0.6}{\hour}) & \SI{28}{\second} \\
 & & & hybrid & 20 ($2\times10$) & GPU & 20.1, \SI{21.4}{\hour} & \SI{2.1}{\hour} \\
a-\ce{TiO2} & 270 & MACE-MPA-0 & random + relax & 100 & GPU & \SI{1.9}{\hour} (\SI{0.5}{\hour}) & \SI{69}{\second} \\
 & & & hybrid & 20 ($2\times10$) & GPU & 17.5, \SI{17.7}{\hour} & \SI{1.8}{\hour} \\
a-\ce{InGaZnO4} & 350 & MACE-MPA-0 & random + relax & 100 & GPU & \SI{1.4}{\hour} (\SI{0.5}{\hour}) & \SI{50}{\second} \\
 & & & hybrid & 20 ($2\times10$) & GPU & 17.4, \SI{22.7}{\hour} & \SI{2.0}{\hour} \\
a-\ce{SiO2} & 648 & MACE-MPA-0 & random + relax & 100 & GPU & \SI{3.1}{\hour} (\SI{2.0}{\hour}) & \SI{1.9}{\minute} \\
 & & & hybrid & 20 ($4\times5$) & GPU & \SIrange{17.6}{18.7}{\hour} & \SI{3.7}{\hour} \\
a-\ce{SiO2} & 648 & BKS & random + relax & 100 ($4\times25$) & CPU & \SIrange{2.1}{2.9}{\hour} & \SI{6}{\minute} \\
 & & & hybrid & 20 ($20\times1$) & CPU & \SIrange{30}{41}{\hour} & \SIrange{30}{41}{\hour} \\
a-\ce{SiO2} & 648 & BKS & random + relax & 100 & GPU & \SI{1.2}{\hour} (\SI{13}{\minute}) & \SI{42}{\second} \\
 & & & hybrid & 20 ($1\times20$) & GPU & \SI{16.6}{\hour} & \SI{50}{\minute} \\
a-As & 2000 & MACE (Liu \textit{et al.}) & random + relax & 5 & GPU & \SI{20}{\minute} (\SI{5}{\minute}) & \SI{4}{\minute} \\
 & & & hybrid & 5 & GPU & \SI{10.3}{\hour} & \SI{2.1}{\hour} \\
a-As & 2000 & MACE-MPA-0 & random + relax & 5 & GPU & \SI{23}{\minute} (\SI{8}{\minute}) & \SI{5}{\minute} \\
 & & & hybrid & 5 ($4+1$) & GPU & \SI{18.2}{\hour} & \SI{3.6}{\hour} \\
a-As & 2000 & both & cubic-cell relax & 5 per set & GPU & \SIrange{2.5}{9}{\minute} & \SI{<2}{\minute} \\
\bottomrule
\end{tabular}}
\end{table}

\section{EXAFS calculations}
\label{sec:exafs}

This section describes the calculations behind
\mainref{fig:exafs}{Figure~5} of the main text. The a-\ce{Ga2O3}
ensemble was regenerated with the structural-benchmark settings;
the a-\ce{InGaZnO4} spectra use a separate, larger ensemble at fixed
volume. These distinctions are detailed in
Sections~\ref{sec:exafs-ga2o3} and \ref{sec:exafs-igzo}.

\subsection{FEFF and Fourier-transform settings (protocol P)}
\label{sec:exafs-settings}

All spectra in \mainref{fig:exafs}{Figure~5} of the main text were calculated with FEFF8.5L, in its open-source Feff8L distribution \cite{feff85exafs}, through the AmorphGen FEFF interface. The runs used the interface as frozen before the v1.0.0 release; the released module writes byte-identical FEFF inputs and gives the same averaged spectra (data archive). Each absorber is the center of a \SI{10}{\angstrom}
cluster cut from the periodic cell, with periodic images. Potentials
are non-self-consistent (Hedin--Lundqvist exchange, \texttt{EXCHANGE 0}),
one per element, with the populations of the cell
(\texttt{-{}-feff-stoichiometry structure}); paths are limited to an
effective half-path length of \SI{8}{\angstrom} (\texttt{RPATH 8}) and four legs
(\texttt{NLEG 4}), with \texttt{CRITERIA 1.0 0.5}; $S_0^2 = 1$ and no
Debye--Waller factor is added, so the spectra
contain only the static disorder of the structures. $\chi(k)$ is
averaged over the sampled sites of each cell, then with equal weight
over cells, and transformed on a \SI{0.05}{\per\angstrom} grid with a Hanning
window from 3 to \SI{12}{\per\angstrom} ($dk = \SI{1}{\per\angstrom}$, 2048 points),
with $k^2$ weighting for a-\ce{Ga2O3} and $k^3$ for a-\ce{InGaZnO4}, as
in the experiments compared.

\subsection{Validation against FEFF reference calculations}
\label{sec:exafs-checks}

Three crystal references distributed with
FEFF8.5L \cite{feff85exafs} were rebuilt from their lattices by
AmorphGen and recalculated: \ce{NiO} (Ni K, rocksalt, $a = \SI{4.1684}{\angstrom}$),
\ce{UO2} (U L$_3$, fluorite, $a = \SI{5.4682}{\angstrom}$), and Cu (Cu K, fcc,
$a = \SI{3.61}{\angstrom}$, self-consistent potentials). Over
$k = \SIrange{2}{18}{\per\angstrom}$, the relative RMS differences between $\chi(k)$ and the corresponding references are \SIlist{0.0012;0.0014;0.0012}{\percent}, respectively. Without the cell populations, FEFF
infers them from the finite cluster, and the \ce{NiO} and \ce{UO2} spectra
differ from the references by \SIlist{2.8;4.6}{\percent}; for this reason, the
amorphous calculations use \texttt{-{}-feff-stoichiometry structure}. The
EXAFS tests in the AmorphGen test suite include these references when
run with \texttt{pytest -{}-run-feff} and an FEFF executable.

\subsection{\texorpdfstring{a-\ce{Ga2O3}}{a-Ga2O3}: ensemble, spectral fit, and energy reference}
\label{sec:exafs-ga2o3}

The 100 random + relax cells of \mainref{tab:oxides}{Table~1} of the main text were regenerated with AmorphGen v1.0.0, seed 2026,
and the settings of \mainref{sec:code}{Section~2.3} of the main text
(\texttt{-{}-retry-mode none}, TorchSim with batches of 10, double
precision, cubic cell, \SI{0.05}{\electronvolt\per\angstrom}). The regenerated ensemble
has a density of \SI{4.773 \pm 0.051}{\gram\per\cubic\centi\metre}, a Ga--O distance of
\SI{1.9331}{\angstrom} (\SI{2.47}{\angstrom} cutoff), and a CN(Ga) of 4.47, as in that table.
The cells are similar to, but not identical to, the original ones: of the
ten original cells distributed with the AmorphGen tutorials, nine are
reproduced to a median atomic displacement below \SI{0.05}{\angstrom}, and the
tenth relaxed into a nearby minimum.

The mean $k^2\chi(k)$ was fitted to the as-deposited film of
Reis \textit{et al.}\ \cite{reis2023} over
$k = \SIrange{2.5}{12}{\per\angstrom}$, with a scale, an offset,
an energy shift $\Delta E_0$, and an extra disorder term
$\Delta\sigma^2$. The normalized squared residual is 0.043,
compared with 0.68 when only the scale and offset are fitted
(\mainref{fig:exafs}{Figure~5a} of the main text).
Section~\ref{sec:exafs-convergence} tests the sensitivity of this fit
to the FEFF settings.

The calculated and measured spectra are referenced to different
energies. The calculated $\chi(k)$ is referenced to the
threshold of the non-self-consistent FEFF potentials, whereas the measured
spectrum is referenced to the edge energy chosen during background removal,
which Reis \textit{et al.}\ do not state. $\Delta E_0$ absorbs this offset and,
since the two are correlated, any phase difference from bond lengths.
To separate them, the $\beta$-\ce{Ga2O3} crystal \cite{ahman1996},
calculated with protocol P (all eight Ga sites of the conventional cell, so
the tetrahedral and octahedral sites carry equal weight), was fitted
in the same way to the annealed film of the same figure, which contains
$\beta$-\ce{Ga2O3} crystallites, over $k = \SIrange{2.5}{10.5}{\per\angstrom}$. This range stops below the two narrow spikes near \SIlist{10.9;11.45}{\per\angstrom}, which match the Bragg-diffraction artifacts Reis \textit{et al.}\ report on this spectrum. The crystal needs $\Delta E_0 = \SI{8.0}{\electronvolt}$ and
$\Delta\sigma^2 = \SI{0.0043}{\angstrom\squared}$, with a normalized squared residual of 0.15 (0.43 if the spikes are included, up to \SI{12}{\per\angstrom}). The structure of the crystal is known, so most of the shift in the amorphous fit comes from the energy references. Over the same range, the 100 cells need \SI{8.7}{\electronvolt} against the annealed film (residual 0.36), and over the full range \SI{11.6}{\electronvolt} against the as-deposited one; the remaining \SI{3}{\electronvolt} therefore lies between the two measured spectra, in their $E_0$ or their first-shell distance, which this check cannot separate.

The extra $\sigma^2$ of the crystal, \SI{0.0043}{\angstrom\squared}, is about the
thermal Ga--O disorder expected at room temperature, since the crystal
calculation contains no disorder beyond its two Ga sites. The static
amorphous model needs only \SI{0.0012}{\angstrom\squared} against the as-deposited film.
If the thermal disorder is similar in both films, the static first-shell
disorder of the model is therefore broader than that of the film by
about \SI{0.003}{\angstrom\squared}, consistent with its mix of four- and five-fold Ga
(\SIlist{55;41}{\percent}). This rests on that assumption, on the imperfect
crystal fit, and on $\Delta\sigma^2$ being correlated with $\Delta E_0$
and absorbing any other difference in shape.

\subsection{Convergence and fit sensitivity}
\label{sec:exafs-convergence}

Two transforms $A$ and $B$ are compared by
$D_R(A,B) = \lVert \mathrm{FT}_A - \mathrm{FT}_B \rVert / \lVert
\mathrm{FT}_B \rVert$, with the norms of the complex transforms taken
over $0 \le R \le \SI{4}{\angstrom}$ (not phase-corrected). The target
was a change below \SI{2}{\percent} under the numerical refinements
and sampling checks of Table~\ref{tab:exafs-convergence}. The criterion
and threshold were fixed before these runs, but after an earlier screen
of 60 Ga sites had shown that the first target, a \SI{2}{\percent}
change in the raw $k^2\chi(k)$ over \SIrange{3}{12}{\per\angstrom}, is
not reached as the path length grows, because the change lies beyond
\SI{4}{\angstrom} in $R$. The window was chosen for that reason. The reduced-site tests that
exceed this threshold motivate the use of every absorber site in the
reported spectra. Relative to protocol P, the refinements
are tighter path filters (\texttt{CRITERIA 0.25 0.125},
F), \texttt{RPATH 10}\,\si{\angstrom} (R), \texttt{NLEG 6} (L), and a \SI{12}{\angstrom}
cluster (C). Site sampling compares a nested subset of sites with all
sites of each cell (S1) and gives the 95th percentile over 1000 random
subsets (S2). Cell sampling (N1) is half the median $D_R$ between two
random halves of the cells, over 1000 splits; for independent cells, it estimates the sampling error of the mean over all cells.

\begin{table}[H]
\centering
\caption{Convergence of the EXAFS transforms, $D_R$ (\si{\percent}) over
$R \le \SI{4}{\angstrom}$. a-\ce{Ga2O3}: all rows on the 100 cells of \mainref{tab:oxides}{Table~1} of
the main text (\num{16000} Ga absorbers; S1 and S2 at 120 of 160 sites per
cell), except C.
a-\ce{InGaZnO4}: F, R, L, and C on 20 cells $\times$ 10
sites per element; S1 and S2 at 40 of 50 sites per cell on 56 cells; N1 on
504 cells.}
\label{tab:exafs-convergence}
\footnotesize
\begin{tabular}{@{}l l r r r r@{}}
\toprule
 & & a-\ce{Ga2O3} & \multicolumn{3}{c}{a-\ce{InGaZnO4}} \\
\cmidrule(lr){3-3} \cmidrule(lr){4-6}
Row & Test & Ga & Zn & Ga & In \\
\midrule
F & path filters & 0.50 & 0.40 & 0.48 & 0.80 \\
R & path length & 0.11 & 0.10 & 0.07 & 0.01 \\
L & scattering order & 0.51 & 0.11 & 0.24 & 0.04 \\
C & cluster radius & 0.00$^c$ & 0.00 & 0.00 & 0.00 \\
S1 & nested sites & 0.42 & 1.94 & 1.60 & 1.87 \\
S2 & site subsets, 95th pct. & 0.72 & 3.81$^a$ & 2.68$^a$ & 3.08$^a$ \\
N1 & cells & 0.99$^b$ & 1.81 & 1.25 & 1.45 \\
\bottomrule
\end{tabular}

\smallskip\noindent\footnotesize $^a$ Above \SI{2}{\percent}, so all 50 sites
of each element are used in every cell, removing the error from
subsampling absorber sites within a cell. $^b$ 100 cells, all 160 Ga sites each (50/50
halves). An earlier 20-cell ensemble gave \SI{2.11}{\percent}.
$^c$ On that 20-cell ensemble, where the \SI{12}{\angstrom} cluster gives the same
spectra as P, since \texttt{RPATH} already limits the paths.
\end{table}

The criterion covers the transform up to \SI{4}{\angstrom}. The full $k$-space
spectrum is not converged: for a-\ce{Ga2O3}, tighter filters and longer
paths change $k^2\chi(k)$ over \SIrange{3}{12}{\per\angstrom} by \SIlist{4.9;7.5}{\percent}
(relative RMS), through features beyond \SI{4}{\angstrom} in $R$. Because the
comparison with experiment is a $k$-space fit, the fit was repeated
with each refinement on the 100 cells (all \num{16000} sites,
Table~\ref{tab:exafs-fit-sensitivity}). The residual changes by less than 0.002 and
$\Delta E_0$ by less than \SI{0.07}{\electronvolt}, but $\Delta\sigma^2$ falls by up
to \SI{0.00019}{\angstrom\squared} (R), about three times its cell-sampling interval.
The fitted $\Delta\sigma^2$ is therefore uncertain to about
\SI{\pm 0.0002}{\angstrom\squared} from the FEFF settings alone.

\begin{table}[H]
\centering
\caption{Fit to the a-\ce{Ga2O3} spectrum of Ref.~\cite{reis2023} with
each FEFF setting (100 cells of \mainref{tab:oxides}{Table~1} of the main text, all 160 Ga
sites per cell).
The brackets give the \SI{95}{\percent} interval of the difference from P over 300
paired bootstrap resamples of the cells. C is not repeated (see
Table~\ref{tab:exafs-convergence}).}
\label{tab:exafs-fit-sensitivity}
\footnotesize
\begin{tabular}{@{}l r r r@{}}
\toprule
Setting & Residual & $\Delta E_0$ (\si{\electronvolt}) & $\Delta\sigma^2$ (\SI{e-4}{\angstrom\squared}) \\
\midrule
P & 0.0431 & 11.56 & 12.2 \\
F & 0.0433 & 11.55 [$-0.013$, $-0.005$] & 11.6 [$-0.59$, $-0.46$] \\
R & 0.0416 & 11.51 [$-0.052$, $-0.041$] & 10.2 [$-2.00$, $-1.84$] \\
L & 0.0421 & 11.61 [$+0.051$, $+0.064$] & 12.0 [$-0.23$, $-0.16$] \\
\bottomrule
\end{tabular}
\end{table}

FEFF8.5L keeps at most \num{15000} paths in its final sum and drops the
rest without a warning; AmorphGen rejects such runs, and none of the
runs here came close (at most \num{10319} paths). The a-\ce{Ga2O3} rows F,
R, and L were first evaluated on the earlier 20-cell ensemble
(\SIlist{0.49;0.11;0.54}{\percent}), whose density is lower
(\SI{4.63}{\gram\per\cubic\centi\metre}). The 100-cell results lead
to the same convergence conclusion.

\subsection{\texorpdfstring{a-\ce{InGaZnO4}}{a-InGaZnO4}: ensemble and density controls}
\label{sec:exafs-igzo}

The 504 a-\ce{InGaZnO4} cells (\ce{In50Ga50Zn50O200}, 350 atoms) were
placed at \SI{6.10}{\gram\per\cubic\centi\metre}
with AmorphGen v1.0.0, using seeds 2027--2035. They were relaxed with
MACE-MPA-0 in double precision to \SI{0.05}{\electronvolt\per\angstrom},
with the cell held fixed. Structures whose placement required a larger
cell or whose relaxation did not converge were
discarded (list in the data archive). This
ensemble is separate from the 100 cells of \mainref{tab:oxides}{Table~1} of the main text,
which were relaxed with a variable cubic cell. At \SI{6.10}{\gram\per\cubic\centi\metre}, the cells are
under about \SI{5}{\giga\pascal} with MACE-MPA-0. Relaxing 56 of them at zero pressure
(to \SI{5.32}{\gram\per\cubic\centi\metre}) changes the complex transform by \SIrange{13}{23}{\percent},
mostly in phase, but moves the first-shell maxima by at most \SI{0.003}{\angstrom}
and leaves the height of the In feature at \SI{3.36}{\angstrom} at 0.093, against
0.092 for the same 56 cells at fixed volume.

Cells at the relaxed density of \mainref{tab:oxides}{Table~1} of the main text were also
tested. After cubic-cell relaxation, a second random + relax ensemble
of 100 cells (\SI{5.39 \pm 0.15}{\gram\per\cubic\centi\metre}) contains 36 cells within
the density range of \SI{5.53 \pm 0.07}{\gram\per\cubic\centi\metre} in that table
(mean \SI{5.530 \pm 0.037}{\gram\per\cubic\centi\metre}). Their In, Ga, and Zn
coordination numbers are 5.31, 4.42, and 4.16, and their M--O distances
are \SIlist{2.209;1.930;2.047}{\angstrom}. Both sets of values match
those in that table. Protocol P on all their Zn, Ga, and In sites gives
phase-corrected first-shell maxima of 1.977 (Zn), 1.874 (Ga), and
\SI{2.128}{\angstrom} (In), within \SI{0.008}{\angstrom} of those of the 504 cells. The In
feature, the mean over \SIrange{3.20}{3.50}{\angstrom}, is 0.099 [0.077--0.132] against
0.093 [0.085--0.103], and the Ga side lobes are unchanged (0.220 and
0.210 at \SIlist{1.35;2.39}{\angstrom}). As in the zero-pressure control, the whole
transforms differ more: $D_R$ between the two means is \SIlist{6.1;15.1;9.0}{\percent} (In, Ga, Zn), against a 36-cell sampling N1 of \SIlist{4.8;3.9;6.7}{\percent}. With N1 above \SI{2}{\percent}, the 504 cells remain the reported
ensemble.

\subsection{\texorpdfstring{a-\ce{InGaZnO4}}{a-InGaZnO4}: comparison with experiment}
\label{sec:exafs-comparison}

\paragraph{Phase-corrected transforms.}
The main comparison is with the transforms of the UHV-sputtered film
with \SI{1}{\percent} \ce{O2} in Fig.~4 of Kumara \textit{et al.}\
\cite{kumara2021}, with $k^3$ weighting and a Hanning
window; the $k$ range is not stated. The transforms are
phase-corrected: Fig.~4 is plotted against $R$, and their text places
the first-shell peaks at 1.97, 1.87, and \SI{2.12}{\angstrom} and
identifies them with the fitted Zn--O, Ga--O, and In--O distances. The
maxima read from that figure are 1.934 (Zn), 1.867 (Ga), and
\SI{2.142}{\angstrom} (In). To match, the windowed $k^3\chi(k)$ was multiplied
by $\exp(-i\varphi(k))$ before the transform, with $\varphi$ the phase
of the shortest metal--O single-scattering path from FEFF8.5L, averaged
over 20 sites per element. This shifts the first shells by $+0.43$
(Zn), $+0.43$ (Ga), and \SI[retain-explicit-plus]{+0.50}{\angstrom} (In). The phase-corrected maxima of asymmetric shells lie below their mean distances: for the 36 cells at the density of \mainref{tab:oxides}{Table~1} (Section~\ref{sec:exafs-igzo}), the maxima of 1.977 (Zn), 1.874 (Ga), and \SI{2.128}{\angstrom} (In) are \SIrange{0.06}{0.08}{\angstrom} below the mean M--O distances of 2.047, 1.930, and \SI{2.209}{\angstrom}. The maxima can therefore agree with experiment while the mean distances of that table are longer than the fitted ones. Because the measured $k$
range is unknown, weak features and side lobes, which depend on the
window, are compared only approximately (Table~\ref{tab:exafs-igzo-features}), and
the Zn feature near \SI{1.4}{\angstrom}, which depends on the background removal
of the experiment, is listed but not interpreted. The intervals cover cell sampling
only. The Ga side lobes at \SIlist{1.35;2.39}{\angstrom} flank the first shell,
and their heights depend strongly on the window, so they are not
discussed in the main text. The measured spectra include thermal
disorder, which the calculations lack. Metal--metal paths near
\SI{3.3}{\angstrom} usually have a larger thermal $\sigma^2$ than the first
metal--O shell, so thermal disorder would most likely lower the
calculated In feature relative to the first shell and widen the gap to
experiment; this was not calculated.

\begin{table}[H]
\centering
\caption{Phase-corrected a-\ce{InGaZnO4} features (504 cells), against
those read from Fig.~4 of Ref.~\cite{kumara2021}. Heights are relative
to the first-shell maximum; brackets give the \SI{95}{\percent} interval from 1000
bootstrap resamples of the cells.}
\label{tab:exafs-igzo-features}
\footnotesize
\begin{tabular}{@{}l l r r@{}}
\toprule
Edge & Feature & Calculated & Measured \\
\midrule
Zn & first-shell maximum (\si{\angstrom}) & 1.969 & 1.934 \\
Ga & first-shell maximum (\si{\angstrom}) & 1.871 & 1.867 \\
In & first-shell maximum (\si{\angstrom}) & 2.128 & 2.142 \\
Ga & side lobe at \SI{1.35}{\angstrom} & 0.224 [0.223--0.225] & 0.303 \\
Ga & side lobe at \SI{2.39}{\angstrom} & 0.210 [0.207--0.213] & 0.282 \\
Zn & feature near \SI{1.35}{\angstrom} & 0.223 [0.221--0.224] at \SI{1.47}{\angstrom} & 0.446 \\
Zn & feature near \SI{3.0}{\angstrom} & 0.127 [0.114--0.141] & 0.12 \\
Ga & feature at \SI{3.10}{\angstrom} & 0.134 [0.124--0.146] & 0.024 (none) \\
In & shoulder near \SI{2.6}{\angstrom} & 0.186 [0.182--0.191] & 0.269 \\
In & local maximum near \SI{3.3}{\angstrom} & 0.112 [0.099--0.126] & 0.236 \\
In & mean over \SIrange{3.20}{3.50}{\angstrom} & 0.093 [0.085--0.103] & 0.208 \\
\bottomrule
\end{tabular}
\end{table}

\paragraph{Uncorrected transforms.}
The transforms of a second film, by Jia \textit{et al.}\
\cite{jia2018}, give an independent check of the phase correction and test the In feature.
They show In, Ga, and Zn K-edge transforms of a film sputtered
from a 1:1:1 target, amorphous as grown and after \SI{1}{\hour} in air at
\SIlist{300;600}{\degreeCelsius}, without phase correction ($k^3$, Hanning window,
$k = \SIrange{2}{10}{\per\angstrom}$ for In and Zn, \SIrange{3}{11}{\per\angstrom} for Ga) and
without $\chi(k)$. We read values from their Fig.~1, to about 0.03 of
the first-shell height and with the lowest point of each curve as zero,
and transformed the mean $\chi(k)$ of the 504 cells with the same weight
and windows ($dk = \SI{1}{\per\angstrom}$ assumed), also without phase
correction. Their first-shell maxima, 1.64 (In), 1.40--1.42 (Ga), and
\SIrange{1.42}{1.53}{\angstrom} (Zn), compare with \SIlist{1.56;1.41;1.47}{\angstrom} calculated.
They lie \SIrange{0.40}{0.51}{\angstrom} below those of Ref.~\cite{kumara2021}, close
to the FEFF shifts above, and the maxima of Ref.~\cite{kumara2021} lie
within \SI{0.04}{\angstrom} of the distances fitted by Jia \textit{et al.}\
(2.13--2.16, 1.87, and \SI{1.97}{\angstrom}), as expected for
phase-corrected transforms. The In feature peaks at \SIrange{2.95}{3.06}{\angstrom} in
these films. Over \SIrange{2.7}{3.0}{\angstrom}, the uncorrected equivalent of the
\SIrange{3.20}{3.50}{\angstrom} window, its mean is 0.17, 0.10, and 0.17 (as grown,
\SIlist{300;600}{\degreeCelsius}), against 0.08 calculated. An energy shift of
\SI{\pm 12}{\electronvolt} moves the calculated value by at most 0.01 and the
calculated In first shell by up to \SI{0.1}{\angstrom}. The calculated
feature is therefore too weak with no phase correction, although the
measured height varies between films. The calculated Ga K-edge lobes
near \SIlist{2.0;2.4}{\angstrom} match those of the as-grown film (0.16 and 0.12
against 0.17 and 0.13), but the first-shell-only fit in Fig.~1(d) of
Ref.~\cite{jia2018} has the same lobes, so they are ripples of the
transform window, not further shells. At the Zn K edge, the noise beyond
\SI{2}{\angstrom}, about 0.15 of the first shell, rules out a comparison, and the
feature near \SI{1}{\angstrom}, the uncorrected counterpart of the one near
\SI{1.4}{\angstrom}, ranges from 0.27 to 0.55 between the films, as expected if it
depends on the background removal. Only values read from the figure are
used.

\paragraph{Fits in reciprocal space.}
Among the experimental datasets compared here, Revenant \textit{et al.}\
\cite{revenant2015} provide measured a-\ce{InGaZnO4} $\chi(k)$ at the Zn and Ga K
edges, for a different film: sol--gel, porous, with In:Ga:Zn =
3:0.4:2 and Ga- and Zn-rich nanoclusters after annealing. We used the data of the film annealed at \SI{450}{\degreeCelsius} from their Figs.~1b and 2b and fitted the mean $k^2\chi(k)$ of the 504 cells as for a-\ce{Ga2O3}, with a scale, an
offset, an energy shift, and an extra $\sigma^2$ over
$k = \SIrange{2.5}{12}{\per\angstrom}$ (\mainref{fig:exafs}{Figure~5c} of the main text). The fits give
energy shifts of 9.9 (Zn) and \SI{13.0}{\electronvolt} (Ga), similar to the \SI{11.6}{\electronvolt} of
a-\ce{Ga2O3}, extra $\sigma^2$ of \SIlist{0.0013;0.0008}{\angstrom\squared}, scales of
0.67 and 0.78, and normalized squared residuals of 0.19 and 0.15,
against 0.73 and 0.91 with the scale and offset only. Because the
composition and structure of this film differ from those of the model,
the comparison tests only the phase and frequency of the first-shell
oscillation. No measured In K-edge $\chi(k)$ is available, so \mainref{fig:exafs}{Figure~5c}
shows the calculated In spectrum without any shift.

\subsection{\texorpdfstring{a-\ce{InGaZnO4}}{a-InGaZnO4}: annealed cells}
\label{sec:exafs-annealed}

We tested whether annealing raises the In feature by applying the hybrid
route of the main text to the 56 cells used for the zero-pressure
control, starting from their fixed-volume structures. The calculations
used AmorphGen v1.0.0, MACE-MPA-0, and single-precision MD with a
\SI{0.5}{\femto\second} timestep. The protocol comprised \SI{50}{\pico\second} at
\SI{3000}{\kelvin}, a quench at \SI{100}{\kelvin\per\pico\second}, and \SI{10}{\pico\second} at \SI{300}{\kelvin}, followed by
relaxation to \SI{0.05}{\electronvolt\per\angstrom}. The cell was held fixed,
so the density remained at \SI{6.10}{\gram\per\cubic\centi\metre}. All 56 relaxations converged, every cell
was liquid at \SI{3000}{\kelvin}, and none has O--O contacts below \SI{2.0}{\angstrom}. The
statistic and decision criteria were specified before the anneal. The
statistic $H$ is the mean of the normalized, phase-corrected
$|\chi(R)|$ over \SIrange{3.20}{3.50}{\angstrom}, and $\Delta H$ is its change between
each annealed cell and the same cell before the anneal, with a \SI{95}{\percent}
interval from 1000 paired bootstrap resamples of the cells. The anneal
strengthens the feature if the lower bound of the interval for $\Delta H$ is above zero,
and causes no change of interest if the whole interval lies within
$\pm 0.06$, half the gap to experiment. With protocol P on all 2800 In
sites, $H$ rises from 0.086 to 0.108 [0.084--0.140], and
$\Delta H = +0.023$ [$-0.018$, $+0.057$]. By the declared margin,
this is no change of interest, although the interval does not exclude
a rise of up to 0.06; $H$ stays well below the measured 0.208. The
anneal does change the spectra as a whole: between the annealed and
the original cells, $D_R$ is \SIlist{16;7;18}{\percent} for Zn, Ga, and In,
mostly from \SIrange{2.0}{3.2}{\angstrom}, outside the window of $H$. For the In
transform of the annealed cells, the refinements F, R, L, and C change
$D_R$ by at most \SI{0.79}{\percent} (20 cells $\times$ 10 sites), as for the
random cells. With $k$ windows of 2--10 and \SIrange{3}{14}{\per\angstrom}, or
without the phase correction (window \SIrange{2.70}{3.00}{\angstrom}), $H$ also rises,
by 0.02--0.03. These checks are not decisive. Between 3.0 and
\SI{3.6}{\angstrom}, each In has slightly more Ga and Zn neighbors after the
anneal (2.53 against 2.35 Ga and 2.44 against 2.31 Zn).

\subsection{Computational cost of EXAFS}
\label{sec:exafs-cost}

The median time per FEFF8.5L run (\SI{10}{\angstrom} cluster, about
300--350 atoms) on one core of the BlueBEAR cluster was
\SI{2.5}{\second} per Ga site in a-\ce{Ga2O3} and \SIlist{3.5;3.4;1.9}{\second} per Zn, Ga,
and In site in a-\ce{InGaZnO4}. The \num{16000} a-\ce{Ga2O3} runs took
11.4 core-hours. For a timing subset of 448 of the 504 a-\ce{InGaZnO4}
cells, the \num{67200} runs took 55 core-hours, about 0.12 core-hours
per cell for all three edges. The independent absorber calculations
can be distributed across CPU cores. The number of cells needed differs between systems
(Table~\ref{tab:exafs-convergence}): the earlier 20 cells of a-\ce{Ga2O3} reach
N1~=~\SI{2.1}{\percent}, while 56 cells of a-\ce{InGaZnO4} give \SIrange{3.9}{5.4}{\percent} and
several hundred are needed for \SI{2}{\percent}. N1 should therefore be checked
for each system.

\bibliographystyle{elsarticle-num}
\bibliography{supplementary}